# Linking Stellar Populations to H II Regions across Nearby Galaxies. II. Infrared Reprocessed and UV Direct Radiation Pressure in H II Regions

Debosmita Pathak[1,2], Adam K. Leroy[1,2], Todd A. Thompson[1,2,3], Laura A. Lopez[1,2], Ashley. T. Barnes[4],
Daniel A. Dale[5], Ian Blackstone[2,3], Simon C. O. Glover[6], Shyam H. Menon[7,8], Jessica Sutter[9],
Thomas G. Williams[10], Dalya Baron[11], Francesco Belfiore[12], Frank Bigiel[13], Alberto D. Bolatto[14],
Médéric Boquien[15], Rupali Chandar[16], Mélanie Chevance[6], Ryan Chown[1], Kathryn Grasha[17,18,28], Brent Groves[19],
Ralf S. Klessen[6,20,21,22], Kathryn Kreckel[23], Jing Li[23], J. Eduardo Méndez-Delgado[23], Erik Rosolowsky[24],
Karin Sandstrom[25], Sumit K. Sarbadhicary[1,2,3], Jiayi Sun[26], and Leonardo Úbeda[27]
[1] Department of Astronomy, Ohio State University, 180 W. 18th Avenue, Columbus, OH 43210, USA; pathak.89@buckeyemail.osu.edu
[2] Center for Cosmology and Astroparticle Physics, 191 West Woodruff Avenue, Columbus, OH 43210, USA
[3] Department of Physics, Ohio State University, 91 West Woodruff Avenue, Columbus, OH 43210, USA
[4] European Southern Observatory, Karl-Schwarzschild-Straße 2, 85748 Garching, Germany
[5] Department of Physics and Astronomy, University of Wyoming, Laramie, WY 82071, USA
[6] Universität Heidelberg, Zentrum für Astronomie, Institut für Theoretische Astrophysik, Albert-Ueberle-Str. 2, 69120 Heidelberg, Germany
[7] Department of Physics and Astronomy, Rutgers University, 136 Frelinghuysen Road, Piscataway, NJ 08854, USA
[8] Center for Computational Astrophysics, Flatiron Institute, 162 5th Avenue, New York, NY 10010, USA
[9] Whitman College, 345 Boyer Avenue, Walla Walla, WA 99362, USA
[10] Sub-department of Astrophysics, Department of Physics, University of Oxford, Keble Road, Oxford OX1 3RH, UK
[11] The Observatories of the Carnegie Institution for Science, 813 Santa Barbara Street, Pasadena, CA 91101, USA
[12] INAF—Osservatorio Astrofisico di Arcetri, Largo E. Fermi 5, I-50125, Florence, Italy
[13] Argelander-Institut für Astronomie, Universität Bonn, Auf dem Hügel 71, 53121 Bonn, Germany
[14] Department of Astronomy and Joint Space-Science Institute, University of Maryland, College Park, MD 20742, USA
[15] Université Côte d'Azur, Observatoire de la Côte d'Azur, CNRS, Laboratoire Lagrange, 06000, Nice, France
[16] Ritter Astrophysical Research Center, University of Toledo, Toledo, OH 43606, USA
[17] Research School of Astronomy and Astrophysics, Australian National University, Canberra, ACT 2611, Australia
[18] ARC Centre of Excellence for All Sky Astrophysics in 3 Dimensions (ASTRO 3D), Australia
[19] International Centre for Radio Astronomy Research, University of Western Australia, 7 Fairway, Crawley, 6009 WA, Australia
[20] Universität Heidelberg, Interdisziplinäres Zentrum für Wissenschaftliches Rechnen, Im Neuenheimer Feld 225, 69120 Heidelberg, Germany
[21] Harvard-Smithsonian Center for Astrophysics, 60 Garden Street, Cambridge, MA 02138, USA
[22] Radcliffe Institute for Advanced Studies at Harvard University, 10 Garden Street, Cambridge, MA 02138, USA
[23] Astronomisches Rechen-Institut, Zentrum für Astronomie der Universität Heidelberg, Mönchhofstraße 12-14, D-69120 Heidelberg, Germany
[24] Department of Physics, University of Alberta, Edmonton, AB T6G 2E1, Canada
[25] Department of Astronomy & Astrophysics, University of California, San Diego, 9500 Gilman Drive, La Jolla, CA 92093, USA
[26] Department of Physics and Astronomy, McMaster University, 1280 Main Street West, Hamilton, ON L8S 4M1, Canada
[27] Space Telescope Science Institute, 3700 San Martin Drive, Baltimore, MD 21218, USA
Received 2024 September 2; revised 2025 January 14; accepted 2025 January 31; published 2025 March 25

## Abstract

Radiation pressure is a key mechanism by which stellar feedback disrupts molecular clouds and drives H II region expansion. This includes direct radiation pressure exerted by UV photons on dust grains, pressure associated with photoionization, and infrared (IR) radiation pressure on grains due to dust-reprocessed IR photons. We present a new method that combines high-resolution mid-IR luminosities from JWST-MIRI, optical attenuation, and nebular line measurements from the Very Large Telecope Multi-Unit Spectroscopic Explorer (VLT-MUSE), and the Hubble Space Telescope (HST) Hα-based region sizes to estimate the strength of radiation pressure in ≈18,000 H II regions across 19 nearby star-forming galaxies. This is the most extensive and direct estimate of these terms beyond the Local Group to date. In the disks of galaxies, we find that the total reprocessed IR pressure is on average 5% of the direct UV radiation pressure. This fraction rises to 10% in galaxy centers. We expect reprocessed IR radiation pressure to dominate over UV radiation pressure in regions where $L_{\rm F2100W}/L_{\rm H\alpha}^{\rm corr} \gtrsim 75$. Radiation pressure due to H ionizations is lower than pressure on dust in our sample, but appears likely to dominate the radiation pressure budget in dwarf galaxies similar to the Small Magellanic Cloud. The contribution from all radiation pressure terms appears to be subdominant compared to thermal pressure from ionized gas, reinforcing the view that radiation pressure is most important in compact, heavily embedded, and young regions.

*Unified Astronomy Thesaurus concepts:* Stellar feedback (1602); H II regions (694); Interstellar medium (847); Extragalactic astronomy (506); Dust physics (2229); Infrared astronomy (786)

*Materials only available in the* online version of record: machine-readable table

---

[28] ARC DECRA Fellow.

## 1. Introduction

Observations show that star formation is an inefficient process. Contrasting the typical volume density of a molecular cloud with the observed gas depletion time suggests that only a small fraction, ∼0.5%, of the total molecular gas mass in galaxies is converted to stars per gravitational freefall time (e.g., D. Utomo et al. 2018; J. Sun et al. 2022, 2023; in nearby galaxies). One popular explanation for this inefficiency is that stellar feedback disperses molecular clouds, thus regulating the rate of star formation (see, e.g., O. Agertz & A. V. Kravtsov 2016; E. C. Ostriker & C.-G. Kim 2022). Here stellar feedback refers to the injection of energy and momentum into the interstellar medium (ISM) by young, massive stars. This feedback has multiple modes and shapes the ISM on multiple length and time scales (see recent reviews by R. S. Klessen & S. C. O. Glover 2016; M. Chevance et al. 2023; E. Schinnerer & A. K. Leroy 2024).

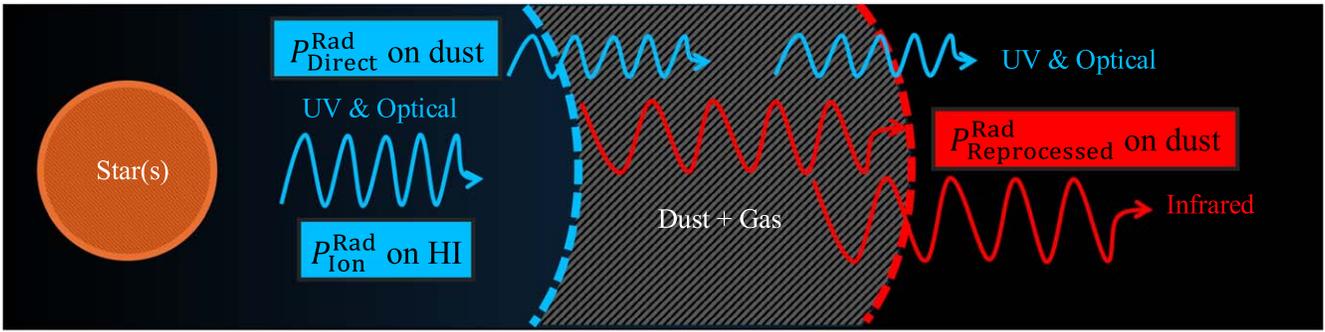

**Figure 1.** Radiation pressure in H II regions. UV and optical photons emitted from the central powering source interact with the surrounding dusty ISM. This results in $P_{\text{Direct}}^{\text{Rad}}$ due to absorption and scattering off of dust grains and $P_{\text{Ion}}^{\text{Rad}}$ due to H ionization. IR-reprocessed photons emitted by dust grains also interact with the dusty ISM. This results in an additional $P_{\text{Reprocessed}}^{\text{Rad}}$ due to absorption of these IR photons by dust. With the assumptions indicated, H$\alpha$ emission is used to estimate $P_{\text{Ion}}^{\text{Rad}}$. H$\alpha$ and the Balmer decrement extinction are used to estimate $P_{\text{Direct}}^{\text{Rad}}$. Mid-IR dust emission and estimated mid-IR extinction are used to calculate $P_{\text{Reprocessed}}^{\text{Rad}}$. See details in Section 3.

Recent results suggest a particularly important role for pre-supernova feedback in dispersing molecular clouds and regulating the star formation efficiency of molecular gas. This is supported by both simulations (e.g., C. D. Matzner 2002; N. Murray et al. 2010; J. E. Dale et al. 2012, 2013; O. Agertz et al. 2013; M. A. Skinner & E. C. Ostriker 2015; S. Raskutti et al. 2016; A. Gatto et al. 2017; D. Rahner et al. 2017; J.-G. Kim et al. 2018, 2021; R. Kannan et al. 2020; S. M. R. Jeffreson et al. 2021; L. Lancaster et al. 2021a; M. Y. Grudić et al. 2022; S. H. Menon et al. 2023; E. P. Andersson et al. 2024) and observations (e.g., N. Murray 2011; K. Grasha et al. 2018, 2019; J. M. D. Kruijssen et al. 2019; M. Chevance et al. 2020; J. K. Barrera-Ballesteros et al. 2021a, 2021b; J. Kim et al. 2021, 2023; M. Chevance et al. 2022; L. Bonne et al. 2023).

Given this importance, studies have examined the impact of different pre-supernova feedback mechanisms in driving the expansion of H II regions in the Galaxy (e.g., M. R. Rugel et al. 2019; E. J. Watkins et al. 2019; A. T. Barnes et al. 2020; G. M. Olivier et al. 2021), the Magellanic Clouds (e.g., M. S. Oey 1996a, 1996b; L. A. Lopez et al. 2011, 2014; M. Chevance et al. 2016; M. Y. Lee et al. 2019; A. F. McLeod et al. 2019), and nearby galaxies (e.g., J. M. D. Kruijssen et al. 2019; A. F. McLeod et al. 2020, 2021; A. T. Barnes et al. 2021, 2022; M. Cosens et al. 2022; L. Della Bruna et al. 2022; I. Blackstone & T. A. Thompson 2023; O. V. Egorov et al. 2023; E. J. Watkins et al. 2023).

This paper focuses on one of the key modes of stellar feedback, radiation pressure, in H II regions, which is schematically illustrated in Figure 1. Young, massive stars output most of their energy and momentum in far-UV (FUV) photons ($\gtrsim 8$ eV; M. R. Krumholz et al. 2019). Most of these are captured by the surrounding ISM via photoionization, line absorption or scattering, and scattering or absorption by dust grains. Photons with energies >13.6 eV can impart momentum to gas directly during photoionization. For lower-energy photons, dust grains have larger scattering and absorption cross sections compared to gas. As a result, most optical–UV photons below the Lyman continuum limit interact with dust grains in metal-rich galaxies like the Milky Way. The momentum transfer of UV photons interacting with dust grains results in "direct" radiation pressure (W. G. Mathews 1967; B. T. Draine 2011a), which also exerts a force on the gas that is dynamically coupled to dust under typical ISM conditions. This radiation pressure can help provide support against gravitational collapse, limit gas accretion onto massive stars, disperse molecular clouds, drive the expansion of H II regions, and even launch local outflows (e.g., N. Z. Scoville et al. 2001; B. A. Groves et al. 2004; M. A. Dopita et al. 2005, 2006; M. R. Krumholz & C. D. Matzner 2009; S. M. Fall et al. 2010; N. Murray et al. 2010; T. Peters et al. 2010; P. F. Hopkins et al. 2011; S. W. Davis et al. 2014; S. Raskutti et al. 2016; R. M. Crocker et al. 2018b; J.-G. Kim et al. 2018; T.-E. Rathjen et al. 2021; S. H. Menon et al. 2023).

Dust re-emits energy from absorbed UV photons at infrared (IR) wavelengths. When absorbed, these "reprocessed" IR photons also deposit momentum into the ISM. In principle, in dusty regions, multiple absorptions can allow this reprocessed radiation pressure to even exceed the direct radiation pressure (e.g., T. A. Thompson et al. 2005; N. Murray et al. 2010; J.-G. Kim et al. 2016; A. K. Leroy et al. 2018; G. M. Olivier et al. 2021). Because reprocessed radiation pressure depends on the optical depth at IR wavelengths, it becomes more effective in regions with high gas column densities, for example, during the early, obscured phase of star formation, in central molecular zones (CMZs), or starburst galaxies (e.g., M. G. Wolfire & J. P. Cassinelli 1987; T. A. Thompson et al. 2005; N. Murray et al. 2010; B. H. Andrews & T. A. Thompson 2011; M. A. Skinner & E. C. Ostriker 2015; R. M. Crocker et al. 2018a; A. K. Leroy et al. 2018; S. Reissl et al. 2018; T.-E. Rathjen et al. 2021; I. Blackstone & T. A. Thompson 2023; S. H. Menon et al. 2023).[29]

In practice, the importance of reprocessed radiation pressure to the disruption of molecular clouds and expansion of H II regions is debated (e.g., S. Reissl et al. 2018; S. H. Menon et al. 2023) because the strength of radiation pressure has been difficult to estimate across a broad range of systems. This calculation requires resolving individual H II regions, which often have radii ≈5–15 pc (S. Sharpless 1959; A. W. Rodgers et al. 1960; L. D. Anderson et al. 2014) as well as estimating the dust opacity and UV, optical, and IR luminosities associated with each region. This requires good multiwavelength coverage at high physical resolution.

Here, we leverage multiwavelength, high-resolution observations from the Physics at High Angular resolution in Nearby Galaxies (PHANGS) surveys (A. K. Leroy et al. 2021; E. Emsellem et al. 2022; J. C. Lee et al. 2022, 2023) to estimate the strength of this feedback term for ∼18,000 H II regions in 19 galaxies. We use new Hubble Space Telescope (HST)-based H II region size estimates (A. T. Barnes et al. 2022; A. Barnes et al. 2025, in preparation; R. Chandar et al. 2025), optical depth and luminosity estimates based on Very Large Telecope Multi-Unit Spectroscopic Explorer (VLT-MUSE) spectroscopy (E. Emsellem et al. 2022; F. Belfiore et al. 2023; B. Groves et al. 2023), and JWST-based IR luminosities. The ability of JWST-MIRI to measure the IR luminosity of individual regions for large samples of galaxies is particularly important. Previous IR telescopes have had poor angular resolution. For example, the ∼6″ resolution of Spitzer at 24 μm corresponds to ≈300 pc at 10 Mpc, making it almost impossible to even isolate individual H II regions outside the Local Group.

This work represents a natural next step and complement to studies of smaller samples of regions in the Milky Way (e.g., A. T. Barnes et al. 2020; G. M. Olivier et al. 2021), Magellanic Clouds (e.g., L. A. Lopez et al. 2011, 2014; E. W. Pellegrini et al. 2011), and the nearest galaxies (e.g., A. F. McLeod et al. 2021; I. Blackstone & T. A. Thompson 2023). Our work also follows directly from A. T. Barnes et al. (2021). They used PHANGS VLT/MUSE Integral Field Unit mapping (E. Emsellem et al. 2022) to estimate the thermal gas pressure and direct radiation pressure for approximately the same set of regions that we study. Since then, A. Barnes et al. (2025, in preparation) produced HST-based region size estimates for many of these regions. As shown by A. T. Barnes et al. (2022), the smaller sizes revealed by HST imply significantly higher radiation pressures than indicated by unresolved seeing-limited observations. The PHANGS-JWST survey (J. C. Lee et al. 2023) has also produced resolved mid-IR maps that yield luminosities for each individual region (e.g., F. Belfiore et al. 2023). This unlocks the ability to make the first robust estimates of all types of radiation pressure for a large sample of regions.

In Section 2, we introduce our multiwavelength data set. In Section 3 we describe our method for calculating radiation pressures and present estimates of the reprocessed IR radiation pressure (Section 3.1), direct optical–UV radiation pressure on dust (Section 3.2), and direct radiation pressure on gas due to ionizing photons (Section 3.3). We then compare the radiation pressure components across different environments (Section 4). In Section 5, we discuss key sources of uncertainty and logical next steps. Section 6 summarizes our key results.

## 2. Data and Methods

We estimate radiation pressures for ≈18,000 H II regions in the 19 galaxies listed in Table 1. This sample represents the overlap of PHANGS-MUSE, PHANGS-JWST, and the HST Hα survey. The specific selection descends from the PHANGS–ALMA pilot study, but broadly the sample consists of massive spiral galaxies (stellar mass $2.5 \times 10^9 M_\odot$–$9.8 \times 10^{10} M_\odot$, star formation rate (SFR) 0.3–16.9 $M_\odot$ yr$^{-1}$) that lie on or near the main sequence of star-forming galaxies. The sample includes 17 barred galaxies and two lower-mass dwarf spirals.

We target H II regions defined by B. Groves et al. (2023) based on VLT-MUSE spectral mapping. We also use the VLT-MUSE data to trace young stellar populations and estimate the local extinction. We estimate IR luminosities of regions from JWST-MIRI imaging. For 17 targets, we use high-resolution narrowband HST Hα imaging to estimate region sizes. For H II without HST size measurements, we estimate sizes from Hα luminosities via a size–luminosity relation defined using HST data (A. Barnes et al. 2025, in preparation; see Appendix A).

### 2.1. PHANGS-MUSE and Nebular Catalogs

We target H II regions drawn from the PHANGS-MUSE nebular catalogs (B. Groves et al. 2023; see also K. Kreckel et al. 2019; F. Santoro et al. 2022). In these catalogs, the footprints of individual H II regions are constructed by applying a modified version of HIIPHOT (D. A. Thilker et al. 2000) to the "convolved and optimized" PHANGS-MUSE Hα emission-line maps (E. Emsellem et al. 2022). We use the Baldwin–Phillips–Terlevich (BPT; J. A. Baldwin et al. 1981) emission-line diagnostics from B. Groves et al. (2023) to select H II regions and remove contaminants such as active galactic nuclei, planetary nebulae, or supernova remnants. This selection is not perfect, e.g., 35% of the supernova remnant candidates identified by J. Li et al. (2024) overlap BPT diagram-selected H II regions. But overall we expect the selection to work well, especially because H II regions represent the most numerous nebulae in our targets (e.g., out of our sample of 20,960 H II regions, only 3% overlap with supernova remnant candidates).

For each H II region, the nebular catalog provides estimates of the extinction-corrected Hα luminosity, $L_{H\alpha}^{\rm corr}$, and the Hα attenuation, $A_{H\alpha}$, determined using the Balmer decrement method assuming a screen model and J. A. Cardelli et al. (1989,

---

[29] Resonantly scattered photons, especially Lyα, can contribute analogously to the "reprocessed" term at low metallicity (e.g., T. Kimm et al. 2018; A. U. Kapoor et al. 2023), but we study mostly massive, metal-rich targets where these photons get absorbed by dust before they have the chance to scatter repeatedly, and hence we do not consider this term here.

Table 1
Target Specifications and Global Properties

| Galaxy | R.A. (deg) | Decl. (deg) | $i$ (deg) | $D$ (Mpc) | Res.[a] (pc) | $\log M_*$ ($\log_{10} M_\odot$) | log SFR ($\log_{10} M_\odot\ \mathrm{yr}^{-1}$) |
|---|---|---|---|---|---|---|---|
| NGC 0628 | 24.17 | 15.78 | 8.9 | 9.8 | 42.9 | 10.3 | 0.2 |
| NGC 1087 | 41.60 | −0.50 | 42.9 | 15.9 | 69.1 | 9.9 | 0.1 |
| NGC 1300 | 49.92 | −19.41 | 31.8 | 19.0 | 82.9 | 10.6 | 0.1 |
| NGC 1365 | 53.40 | −36.14 | 55.4 | 19.6 | 85.3 | 11.0 | 1.2 |
| NGC 1385 | 54.37 | −24.50 | 44.0 | 17.2 | 75.1 | 10.0 | 0.3 |
| NGC 1433 | 55.51 | −47.22 | 28.6 | 18.6 | 81.3 | 10.9 | 0.1 |
| NGC 1512 | 60.98 | −43.35 | 42.5 | 18.8 | 82.2 | 10.7 | 0.1 |
| NGC 1566 | 65.00 | −54.94 | 29.5 | 17.7 | 77.2 | 10.8 | 0.7 |
| NGC 1672 | 71.43 | −59.25 | 42.6 | 19.4 | 84.6 | 10.7 | 0.9 |
| NGC 2835[b] | 139.47 | −22.35 | 41.3 | 12.2 | 53.4 | 10.0 | 0.1 |
| NGC 3351 | 160.99 | 11.70 | 45.1 | 10.0 | 43.4 | 10.4 | 0.1 |
| NGC 3627 | 170.06 | 12.99 | 57.3 | 11.3 | 49.1 | 10.8 | 0.6 |
| NGC 4254 | 184.71 | 14.42 | 34.4 | 13.1 | 57.2 | 10.4 | 0.5 |
| NGC 4303 | 185.48 | 4.47 | 23.5 | 17.0 | 74.1 | 10.5 | 0.7 |
| NGC 4321 | 185.73 | 15.82 | 38.5 | 15.2 | 66.4 | 10.8 | 0.6 |
| NGC 4535[b] | 188.58 | 8.20 | 44.7 | 15.8 | 68.8 | 10.5 | 0.3 |
| NGC 5068[c] | 199.73 | −21.04 | 35.7 | 5.2 | 22.6 | 9.4 | −0.6 |
| NGC 7496 | 347.45 | −43.43 | 35.9 | 18.7 | 81.6 | 10.0 | 0.3 |
| IC 5332[c] | 353.61 | −36.10 | 26.9 | 9.0 | 39.3 | 9.7 | −0.4 |

**Notes.** Central R.A., decl., and inclination from P. Lang et al. (2020). Distances are from G. S. Anand et al. (2021). Stellar masses and star formation rates (SFRs) are from A. K. Leroy et al. (2021).
[a] Physical resolution corresponding to the 0″.9 resolution of JWST-MIRI at F2100W at the distance to the galaxy.
[b] HST H$\alpha$ observations (R. Chandar et al. 2025) still pending.
[c] Dwarf galaxies that are atomic gas dominated, lower metallicity, with lower ISM pressure.

CCM89) extinction law (see also F. Belfiore et al. 2023; also note that we translate back to $A_{\mathrm{H}\alpha}$ from the catalog-provided $A_V$). We use $L_{\mathrm{H}\alpha}^{\mathrm{corr}}$ to infer the bolometric luminosity of the powering stellar population and $A_{\mathrm{H}\alpha}$ to infer the dust attenuation in the IR and UV. Requiring reliable $A_{\mathrm{H}\alpha}$ measurements yields a sample of 20,683 regions. Restricting our analysis to H II regions that have full JWST-MIRI coverage (i.e., excluding regions at the edge of MIRI mosaics), we arrive at a final sample of 17,615 H II regions.

In areas of high H$\alpha$ surface brightness, the H II region boundaries in the B. Groves et al. (2023) catalog become unreliable and often blend smaller regions together. This inflates the size and luminosity of these regions. This is an issue in galactic centers, which in some galaxies contribute only a handful of H II regions despite containing a large fraction of the total H$\alpha$ emission. By eye, many of the H II regions identified in galaxy centers indeed appear to be a blend of multiple sources. For example, in NGC 4303, the entire galaxy center is identified as two large H II regions, while in NGC 4535, the entire galaxy center is counted as a single H II region.

To address this, in galaxy centers we also calculate the radiation pressure for each individual pixel within H II regions (similar to L. A. Lopez et al. 2011, spatial mapping of pressures in 30 Doradus). To do this, we use the PHANGS-MUSE maps of attenuation-corrected H$\alpha$ intensity, $A_{\mathrm{H}\alpha}$, and 0″.9 maps of mid-IR emission. These "pixel-by-pixel" measurements would be appropriate in the case where each H II region has size approximately matched to that of our PSF, FWHM ∼ 20–85 pc. However, this is still larger than the typical size of H II regions, even in the luminous star-forming regions in galaxy centers (e.g., E. Schinnerer et al. 2023; B. C. Whitmore et al. 2023; J. Sun et al. 2024). We therefore expect that the size used in this calculation represents an upper limit on the true region size, and therefore our pixel-wise calculations yield lower limits on the pressures.

### 2.2. PHANGS-JWST

We use imaging data from the PHANGS-JWST Cycle-1 Treasury program (GO 2107, PI Lee; J. C. Lee et al. 2023). We calculate the photometric intensities for each H II region in four JWST-MIRI filters: F770W, F1000W, F1130W, and F2100W, centered at 7.7, 10, 11.3, and 21 $\mu$m, respectively. This combination of filters traces the emission from polycyclic aromatic hydrocarbon (PAH) complexes and the dust continuum, with F770W and F1130W more PAH-dominated (e.g., J. D. T. Smith et al. 2007), and F2100W more continuum-dominated (e.g., B. T. Draine & A. Li 2007). The nature of the 10$\mu$m emission remains more ambiguous. The filter is sensitive to silicate absorption (e.g., L. Hao et al. 2005) and continuum emission from small grains (B. T. Draine & A. Li 2007). It may also trace the extended wings of the PAH vibrational bands, based on the observation that F1000W emission in PHANGS-JWST correlates well with the PAH-dominated F770W and F1130W bands (e.g., A. K. Leroy et al. 2023; K. M. Sandstrom et al. 2023).

The PHANGS-JWST observations are described in J. C. Lee et al. (2023), and the data processing is presented by T. G. Williams et al. (2024), to which we refer the reader for details on these data. Briefly, the data are calibrated using a version of the STScI JWST pipeline[30] modified to handle extended sources ("pjpipe"[31]). The JWST-MIRI calibration scheme yields an intensity scale accurate to better than 5% (J. R. Rigby et al. 2023). The MIRI mosaics are astrometrically aligned with paired NIRCam observations using bright point sources. Those NIRCam observations are in turn registered against HST imaging that has been aligned to Gaia DR3 sources. As a result, the overall astrometric accuracy of the images is significantly better than the FWHM of the JWST-MIRI PSF (T. G. Williams et al. 2024). The overall background levels of the MIRI images are set using previous mid-IR imaging and validated to be accurate to better than ±0.1 MJy sr$^{-1}$ (A. K. Leroy et al. 2023; J. C. Lee et al. 2023; T. G. Williams et al. 2024). The H II regions that we study appear bright in all bands in the MIRI images (D. Pathak et al. 2024), and we detect essentially all of them at high signal-to-noise (SNR; F. Belfiore et al. 2023; D. Pathak et al. 2024), making the uncertainty in the flux calibration and background level the most relevant sources of uncertainty. Ninety-seven percent of our 17,615 regions are detected at >20σ in at least one MIRI filter.

---
[30] https://jwst-pipeline.readthedocs.io
[31] https://pjpipe.readthedocs.io

We use the MIRI images convolved to a Gaussian PSF with FWHM of 0″.9. This is the smallest "safe" Gaussian PSF to which we can convolve the F2100W data (native PSF FWHM 0″.674), and the convolution improves the SNR of the images significantly T. G. Williams et al. (2024). Table 1 reports the corresponding physical resolution of the convolved JWST-MIRI imaging. Note that the B. Groves et al. (2023) catalogs and our own additional analysis use the "convolved and optimized" VLT-MUSE Hα maps (E. Emsellem et al. 2022), which have a median angular resolution of 0″.92 (range 0″.56−1″.25). Thus, we treat the PHANGS-JWST data as effectively resolution matched to the PHANGS-MUSE maps.

### 2.3. PHANGS-HST Hα Imaging and Nebular Catalogs

Finally, we use complementary narrowband Hα data from HST (P.I. R. Chandar; R. Chandar et al. 2025) for 17 of our 19 targets. These images are diffraction limited with an FWHM PSF ≈0″.1, corresponding to ≈2–9 pc physical resolution. This is much sharper than the 20–85 pc resolution of the PHANGS-MUSE data. A. Barnes et al. (2025, in preparation) used these HST data to measure sizes for the B. Groves et al. (2023) H II regions, providing a value-added catalog of resolved sizes and refined luminosities that augment the spectroscopic measurements produced by F. Santoro et al. (2022), published in B. Groves et al. (2023), with updated auroral line fits (and associated properties) from M. Brazzini et al. (2024). They identify H II region boundaries within each individual B. Groves et al. (2023) region via surface brightness cuts and then provide size and luminosity estimates that reflect only the smaller HST-visible region. Whenever available, we use their "circular" radii, which translate the area of the region at the isophotal boundary into an equivalent radius via $r = \sqrt{A/\pi}$. We also adopt the $L_{\mathrm{H}\alpha}$ measured from the HST region, which is smaller than the MUSE value.

The HST narrowband Hα data have poorer surface brightness sensitivity and somewhat less areal coverage than the VLT-MUSE data; additionally, catalogs for two of our targets were not available at the time of publication. As a result, only 7082 of the 17,615 selected B. Groves et al. (2023) regions were detected in the HST maps. For regions without HST measurements, we use the MUSE Hα luminosity to predict the size and luminosity that HST would be expected to measure following the procedures detailed in Appendix A. We leverage two tight (<0.5 dex scatter) scaling relations from Barnes et al. (2025, in preparation), one linking the HST and MUSE-derived Hα luminosities and the other linking the HST Hα luminosity to the measured size. We show both relations in Appendix A. Extrapolating in this way allows us to use all of the selected B. Groves et al. (2023) regions, improving the statistical power of the results and expanding the analysis to include lower-luminosity regions often not detected by HST.

Our calculations require estimates of the bolometric luminosity, $L_{\mathrm{bol}}$, for each region. We infer this from the extinction-corrected Hα luminosity, $L_{\mathrm{H}\alpha}^{\mathrm{corr}}$. Similar to A. T. Barnes et al. (2021), we perform a luminosity-weighted average over the first 4 Myr of simple stellar population (SSP) evolution to calculate a typical ratio of $L_{\mathrm{bol}}/L_{\mathrm{H}\alpha}^{\mathrm{corr}}$. Specifically, we use STARBURST99 SSP models (C. Leitherer et al. 1999; C. Leitherer et al. 2014) with the default parameters and a fully populated P. Kroupa (2001) initial mass function (IMF) with a maximum stellar mass of 100 $M_\odot$. This yields $L_{\mathrm{bol}}/L_{\mathrm{H}\alpha}^{\mathrm{corr}} \approx 88$. For reference, literature versions of this ratio span from ≈60 near the zero-age main sequence (ZAMS) to ≈140 for continuous star formation (e.g., see E. Schinnerer & A. K. Leroy 2024). We discuss subtleties related to this calculation in Section 5.

We provide a value-added catalog that complements the MUSE nebular catalog presented in B. Groves et al. (2023) and A. Barnes et al. (2025, in preparation) with the measurements needed to reproduce our work: region sizes, dust attenuations, optical depths, luminosities, and resulting pressure estimates. Table 3 summarizes this data, and Table 4 provides a complete list of all columns contained in our value-added catalog of 17,615 H II regions.

### 3. Radiation Pressure in H II Regions

Radiation pressure scales with the energy per area captured by the ISM. Quantifying $P^{\mathrm{Rad}}$ thus involves estimating the relevant luminosity and size for each H II region, then estimating what fraction of this luminosity is captured by the ISM. For reprocessed radiation pressure, the IR luminosity is relevant. Radiation pressure due to ionizations comes from photons with energy above 13.6 eV. For direct radiation pressure on dust, the optical and nonionizing UV radiation is most relevant, which will be a substantial fraction of the bolometric luminosity for a young star cluster.

For a spherical region with luminosity $L(\lambda)$ and radius $R_{\mathrm{circ}}$, the local flux is

$$F_{\mathrm{local}}(\lambda) = \mathcal{K} \frac{L(\lambda)}{4\pi R_{\mathrm{circ}}^2}, \quad (1)$$

where $\mathcal{K}$ is a geometric factor that depends on where the energy is captured by the ISM. For a shell of ISM material where the luminosity is concentrated within some $r \ll R_{\mathrm{circ}}$,

$$F_{\mathrm{local,shell}}(\lambda) = \frac{L(\lambda)}{4\pi R_{\mathrm{circ}}^2}. \quad (\mathcal{K} = 1) \quad (2)$$

For absorbing ISM material uniformly distributed throughout the region, the volume-averaged local flux is

$$\begin{aligned} F_{\mathrm{local,vol}}(\lambda) &= \frac{1}{V} \int \frac{L(\lambda)}{4\pi r^2} dV \\ &= \frac{3}{4\pi R_{\mathrm{circ}}^3} \int_0^{R_{\mathrm{circ}}} \frac{L(\lambda)}{4\pi r^2} 4\pi r^2 dr \\ &= 3 \frac{L(\lambda)}{4\pi R_{\mathrm{circ}}^2}. \quad (\mathcal{K} = 3) \end{aligned} \quad (3)$$

In principle, the simplest realistic geometric assumption for H II regions is a shell, and all radiation pressure estimates presented in this paper assume $\mathcal{K} = 1$. Uniform-distribution pressures can be derived by simply multiplying all pressures by 3. We discuss these geometric considerations further in Section 5.2.

Finally, we estimate what fraction of the flux imparts momentum to the surrounding ISM by accounting for $\tau(\lambda)$, the optical depth of the shell at wavelength $\lambda$:

$$P^{\rm Rad}(\lambda) = (1 - e^{-\tau(\lambda)}) \frac{F_{\rm local,shell}(\lambda)}{c}$$
$$= (1 - e^{-\tau(\lambda)}) \frac{L(\lambda)}{4\pi R_{\rm circ}^2 c}. \quad (4)$$

When considering pressure due to H ionizations, we only consider absorption. For a dusty shell, $\tau(\lambda)$ will have contributions from both absorption and scattering (e.g., S. Reissl et al. 2018). For absorption, the full momentum of the incident photon contributes. For scattering, the momentum contribution is the product of the incident photon momentum and $(1 - \cos\theta)$, where $\theta$ is the scattering angle (G. B. Rybicki & A. P. Lightman 1985). Averaged over many events, this becomes $(1 - \langle\cos\theta\rangle)$ times the photon momentum, where $\langle\cos\theta\rangle$ will be a function of the dust grain size and $\lambda$. In general, momentum transfer due to scattering is negligible for $\lambda \gtrsim 1\,\mu$m because the cross section for scattering drops rapidly past the near-IR (e.g., S. Reissl et al. 2018). Therefore, we ignore scattering for the IR-reprocessed pressure, and account for scattering when calculating direct UV-optical radiation pressure on dust.

Note that since $L(\lambda) \propto D^2$ and $R_{\rm circ}^2 \propto D^2$ for a fixed angular size, where $D$ is the distance to a target, the expression for $P^{\rm Rad}(\lambda)$ in Equation (4) is independent of the distance as long as the region has been resolved. Also note that in regions with high IR optical depth, IR photons can encounter multiple dusty shells, where Equation (4) applies for each shell, effectively increasing the reprocessed radiation pressure by a factor that will approach $\tau$. Our target H II regions are optically thin to IR photons, so we neglect this effect and only consider a single dusty shell for both the reprocessed and direct terms.

### 3.1. Reprocessed Radiation Pressure

Following Equation (4), estimating $P^{\rm Rad}_{\rm Reprocessed}$ requires the IR luminosity, size, and IR optical depth for each H II region. The H II regions in our sample are optically thin to IR photons (Table 3), so we can approximate $(1 - e^{-\tau}) \to \tau$. Then, the expression for reprocessed radiation pressure becomes an integral over the full IR spectrum ($\lambda \approx 1$–1000 $\mu$m)

$$P^{\rm Rad}_{\rm Reprocessed} = \int_{1\,\mu\rm m}^{1000\,\mu\rm m} \tau(\lambda) \frac{L_{\lambda,\rm dust}(\lambda)}{4\pi R_{\rm circ}^2 c} d\lambda, \quad (5)$$

where $L_{\lambda,\rm dust}$ is the wavelength-specific IR luminosity from dust.[32]

JWST-MIRI provides mid-IR maps at high-enough resolution to isolate individual H II regions from the surrounding diffuse ISM (D. Pathak et al. 2024). However, we lack access to longer wavelengths ($\lambda \gtrsim 25\,\mu$m) at high resolution. To account for the impact of radiation at these longer wavelengths, we first calculate the radiation pressure contribution associated with each of the four MIRI filters—F770W, F1000W, F1130W, and F2100W. Then, we use dust SED models to scale from the mid-IR to total IR radiation pressure. For each MIRI filter FX, the pressure $P^{\rm Rad}_{\rm Reprocessed,FX}$ is

$$P^{\rm Rad}_{\rm Reprocessed,FX} = \tau_{\rm FX} \frac{L_{\rm FX}}{4\pi R_{\rm circ}^2 c}, \quad (6)$$

where $L_{\rm FX} \approx \Delta\lambda_{\rm FX} L_{\rm FX,\lambda}$ is the integrated luminosity in that filter. Formally, $\tau_{\rm FX}$ should be the luminosity-weighted average optical depth over the filter. Since the optical depth does not vary significantly over the narrow width of the MIRI filters, we take $\tau_{\rm FX}$ to be the optical depth at the central wavelength of the filter.

To constrain the optical depth $\tau_{\rm FX}$ in each filter, we extrapolate from the H$\alpha$ attenuation of each region, $A_{\rm H\alpha}$, measured by B. Groves et al. (2023). Formally, these Balmer-decrement-based $A_{\rm H\alpha}$ reflect the attenuation along the whole line of sight. However, the H$\alpha$ and H$\beta$ used in the calculation are localized in H II regions, and we expect that the measured opacity is concentrated near the H II region. To convert from H$\alpha$ to other wavelengths, we use the extinction law of K. D. Gordon et al. (2023, hereafter G23), which is based on multiwavelength observations of interstellar extinction along multiple sight lines and extends out to $\sim 30\,\mu$m. Beyond $\lambda = 30\,\mu$m, we add a continuous transition to a classical power-law extinction curve with $\tau \propto \lambda^\beta$ and $\beta = -2$ at $\lambda \gtrsim 30\,\mu$m (B. T. Draine 2011b). Table 2 summarizes relevant conversions derived from the G23 extinction curve and the B. Groves et al. (2023) $A_{\rm H\alpha}$ measurements. We discuss potential issues and next steps related to the extinction curve in Section 5.3.

In principle, an alternative approach would be to model the full IR spectral energy distribution (SED) and simultaneously constrain the radiation field and dust optical depth (e.g., B. T. Draine & A. Li 2007). Unfortunately, for targets beyond the Magellanic Clouds, the physical resolution of far-IR imaging that captures the $\lambda \gtrsim 100\,\mu$m peak of the SED remains far coarser than the scale of individual H II regions. We do note that lower-resolution observations show a good correlation between $A_{\rm H\alpha}$ and dust extinction derived from IR SED modeling (e.g., K. Kreckel et al. 2013).

Figures 2 and 3 show the result of these first two parts of the calculation and Table 3 reports related numbers. Figure 2 shows the distribution of mid-IR luminosities $L_{\rm FX} = \Delta\lambda_{\rm FX} L_{\rm FX,\lambda}$ for our target regions. Our regions show typical luminosities $\sim 10^{38}$ erg s$^{-1}$ in the continuum-tracing F1000W and F2100W filters, and $\sim 5 \times 10^{38}$–$10^{39}$ erg s$^{-1}$ in the PAH-dominated F770W and F1130W filters. The mid-IR luminosities are $\approx 0.5$–1 dex lower than the median bolometric luminosity of regions (see Section 3.2), indicating that a significant fraction of the bolometric luminosity emerges in the mid-IR (see Section 5.4).

Figure 3 and Tables 2 and 3 show the distribution of $A_{\rm H\alpha}$ for our target regions. Treating all regions equally, the median $A_{\rm H\alpha} \approx 0.63$ mag, with a $\approx 0.7$ mag 16th–84th percentile range. The median $A_{\rm H\alpha}$ rises to $\approx 1.1$ mag if we instead weight by H$\alpha$ luminosity. The mid-IR optical depths are lower than $\tau_{\rm H\alpha}$ by factors of 20–30, with median $\tau_{\rm FX}$ of 0.02 (7.7 $\mu$m), 0.06 (10 $\mu$m), 0.04 (11.3 $\mu$m), and 0.03 (21 $\mu$m) and a 16th–84th percentile range of 0.03. The optical depth falls rapidly beyond the mid-IR, so our regions are optically thin to IR photons.

Figure 4 shows the distribution of $P^{\rm Rad}_{\rm Reprocessed,FX}/k_B$ for each MIRI filter, which are of the order of 10–1000 K cm$^{-3}$. These estimates only capture the IR radiation pressure exerted by photons within the wavelength range of the filter. To estimate

---
[32] For reference, in the limit of large IR $\tau \gg 1$, the relevant optical depth is the Rosseland mean optical depth $\tau_{\rm Rs}$, and photons diffuse outward through multiple dusty shells as described above. This yields $P^{\rm Rad}_{\rm Reprocessed} = \tau_{\rm Rs} L_{\rm bol}/(4\pi R_{\rm circ}^2 c)$.

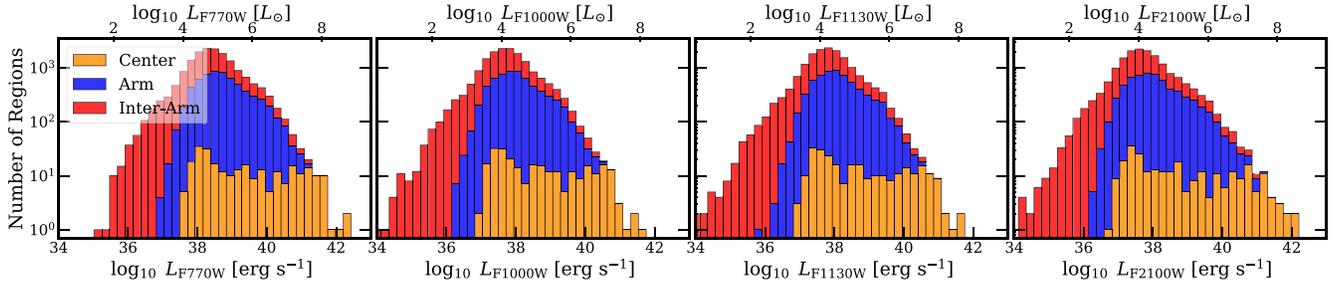

**Figure 2.** The distribution of filter-integrated luminosities $L_{FX} = \Delta\lambda_{FX} L_{\lambda,FX}$ for our sample of H II regions. We separate the regions by environment following M. Querejeta et al. (2021) and plot distributions for galactic centers (yellow), spiral arms and bars (blue), and inter-arm regions (red) separately. The panels show results for the four JWST-MIRI filters (FX) that we use—F770W, F1000W, F1130W, and F2100W.

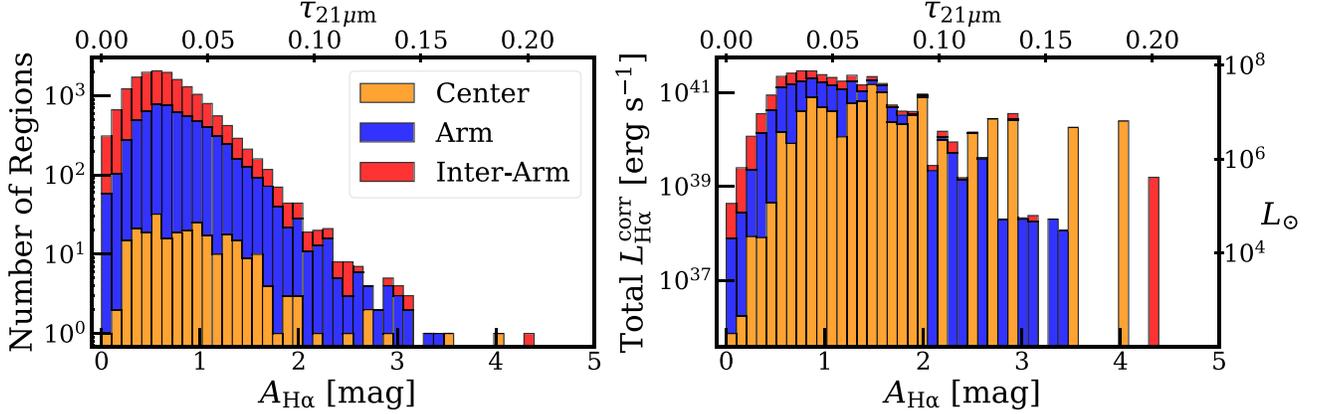

**Figure 3.** Left panel: the distribution of H$\alpha$ attenuation, $A_{H\alpha}$, from VLT-MUSE (Section 2.1) in our sample of H II regions, colored by local environment as in Figure 2. The corresponding optical depth at 21 $\mu$m, $\tau_{21\mu m}$, converted from $A_{H\alpha}$ using the G23 extinction law (Table 2), is included as an alternate $x$-axis. Right panel: corresponding $L_{H\alpha}^{corr}$-weighted histograms, capturing how the H$\alpha$ luminosity is distributed across regions of different attenuation.

**Table 2**
Observed $A_{H\alpha}$ and Relevant G23 Extinction Law Conversions

| Median $A_{H\alpha}$ (1) | $L_{H\alpha}^{corr}$-weighted Median $A_{H\alpha}$ (2) | Median $A_V$ (3) | $\frac{A_{H\alpha}}{A_V}$ (4) | $\frac{A_{7.7\mu m}}{A_V}$ (5) | $\frac{A_{10\mu m}}{A_V}$ (6) | $\frac{A_{11.3\mu m}}{A_V}$ (7) | $\frac{A_{21\mu m}}{A_V}$ (8) | $\frac{A_{150nm}}{A_V}$ (9) |
|---|---|---|---|---|---|---|---|---|
| $0.63_{0.34}^{1.05}$ mag | $1.06_{0.66}^{1.60}$ mag | $0.80_{0.43}^{1.33}$ mag | 0.790 | 0.028 | 0.082 | 0.048 | 0.040 | 2.782 |

**Note.** Balmer-decrement-based attenuation, in magnitudes from B. Groves et al. (2023) for our targets sample of H II regions (Section 2.1). Columns report: (1) median and 16th–84th percentile range of $A_{H\alpha}$ treating all regions equally; (2) median and 16th–84th percentile range of $A_{H\alpha}$ weighting regions by $L_{H\alpha}^{corr}$; (3) corresponding median and 16th–84th percentile range in V-band attenuation $A_V$ converting $A_{H\alpha}$ using the K. D. Gordon et al. (2023, G23) extinction law; and (4)–(9) multiplicative factors from G23 to convert between $A_{H\alpha}$ and attenuation at the V band, 7.7 $\mu$m, 10 $\mu$m, 11.3 $\mu$m, 21 $\mu$m, and our fiducial UV wavelength, $\lambda$ = 150 nm. All conversions assume the Milky Way $R_V = 3.1$ model. The corresponding optical depths can be derived as $\tau_\lambda = A_\lambda/1.086$.

the fraction of the total $P_{Reprocessed}^{Rad}$ captured by the MIRI filters, we combine the G23 extinction curves with the B. S. Hensley & B. T. Draine (2023, hereafter HD23) dust SED models. Appendix B details the full procedure. Briefly, we generate HD23 models for a range of interstellar radiation field (ISRF) intensities. Following B. T. Draine & A. Li (2007), we consider scaled versions of the J. S. Mathis et al. (1983) radiation field, with the intensity indicated by $U$, and $U = 1$ the nominal solar neighborhood ISRF. We consider values from $U = 1$ to $U = 10^5$, an approximate upper bound for plausible ISRF strength in the H II regions in our sample ($U_{max} \approx 10^5$ in 30 Doradus; e.g., M. Y. Lee et al. 2019).

In Appendix B, we show that for a wide range of $U = 1-10^4$, the four MIRI filters together trace a roughly constant fraction, $\approx 30\%$, of the total IR $P_{Reprocessed}^{Rad}$. This translates to a mid-IR-to-total $P_{Reprocessed}^{Rad}$ conversion factor of $f_{MIRI}^{TIR} = 2.7 - 3.7$. We adopt a conversion factor of $f_{MIRI}^{TIR} = 3.33$, with uncertainty of $\lesssim 15\%$ across the plausible range of $U$. For each region, we thus calculate

$$P_{Reprocessed}^{Rad} = f_{MIRI}^{TIR} \sum_{MIRI\ FX} P_{Reprocessed,FX}^{Rad}. \quad (7)$$

In the top row of Figure 5 and Table 3 we summarize the distribution of total IR $P_{Reprocessed}^{Rad}$. We find median $P_{Reprocessed}^{Rad}/k_B$ of the order of 1000 K cm$^{-3}$. The $L_{H\alpha}^{corr}$-weighted histograms shift to higher median $P_{Reprocessed}^{Rad}/k_B \approx$ 1500 K cm$^{-3}$, highlighting how the subset of H II regions with the highest $L_{H\alpha}^{corr}$ also has high $P_{Reprocessed}^{Rad}$. Using the spiral arm and galaxy center masks of M. Querejeta et al. (2021), we check

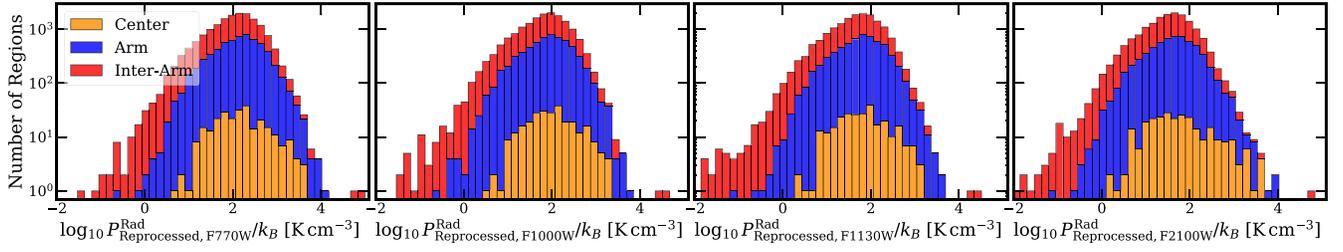

**Figure 4.** Distribution of reprocessed radiation pressures $P_{\text{Reprocessed,FX}}^{\text{Rad}}$ for each individual MIRI filter FX. We separate regions by local environment as in Figure 2. For the values of $P_{\text{Reprocessed}}^{\text{Rad}}$ resulting from combining all filters and applying the SED-based correction described in Appendix B, see Figure 5.

**Table 3**
Summary of H II Region Properties and Pressures

| Quantity | Unit | Unweighted | | | | $L_{\text{H}\alpha}^{\text{corr}}$-weighted | | | |
|---|---|---|---|---|---|---|---|---|---|
| | | All | Center | Arm | Inter-Arm | All | Center | Arm | Inter-Arm |
| $R_{\text{circ}}$ | pc | $8.34_{4.32}^{33.8}$ | $37.4_{4.92}^{120.04}$ | $11.18_{5.1}^{44.8}$ | $7.06_{3.9}^{23.28}$ | $116.77_{48.17}^{204.17}$ | $169.33_{114.67}^{277.42}$ | $96.68_{41.58}^{161.53}$ | $81.08_{30.14}^{164.87}$ |
| $L_{\text{H}\alpha}^{\text{corr}}$ | log $L_\odot$ | $3.47_{2.76}^{4.45}$ | $4.25_{2.77}^{6.08}$ | $3.69_{2.96}^{4.72}$ | $3.31_{2.64}^{4.14}$ | $5.86_{4.95}^{6.75}$ | $6.75_{6.15}^{7.27}$ | $5.58_{4.82}^{6.19}$ | $5.47_{4.49}^{6.16}$ |
| $L_{\text{bol}}$ | log $L_\odot$ | $5.42_{4.7}^{6.39}$ | $6.19_{4.71}^{8.02}$ | $5.63_{4.91}^{6.66}$ | $5.25_{4.58}^{6.09}$ | $7.8_{6.89}^{8.7}$ | $8.7_{8.1}^{9.21}$ | $7.52_{6.76}^{8.14}$ | $7.41_{6.44}^{8.1}$ |
| $L_{\text{F770W}}$ | log $L_\odot$ | $4.87_{4.22}^{5.65}$ | $5.57_{4.49}^{6.68}$ | $5.08_{4.49}^{5.91}$ | $4.74_{4.03}^{5.36}$ | $6.89_{6.02}^{7.96}$ | $7.96_{7.39}^{8.68}$ | $6.62_{5.91}^{7.25}$ | $6.46_{5.53}^{7.16}$ |
| $L_{\text{F1000W}}$ | log $L_\odot$ | $4.18_{3.53}^{4.95}$ | $4.88_{3.87}^{6.74}$ | $4.38_{3.78}^{5.19}$ | $4.02_{3.33}^{4.67}$ | $6.17_{5.3}^{7.19}$ | $7.23_{6.68}^{7.89}$ | $5.89_{5.19}^{6.5}$ | $5.75_{4.81}^{6.45}$ |
| $L_{\text{F1130W}}$ | log $L_\odot$ | $4.27_{3.62}^{5.03}$ | $4.95_{3.89}^{6.79}$ | $4.47_{3.88}^{5.27}$ | $4.11_{3.43}^{4.75}$ | $6.23_{5.37}^{7.31}$ | $7.31_{6.75}^{7.99}$ | $5.96_{5.26}^{6.56}$ | $5.8_{4.89}^{6.52}$ |
| $L_{\text{F2100W}}$ | log $L_\odot$ | $4.15_{3.49}^{5.03}$ | $4.93_{3.75}^{7.06}$ | $4.38_{3.73}^{5.3}$ | $3.99_{3.31}^{4.7}$ | $6.51_{5.44}^{7.66}$ | $7.78_{7.06}^{8.44}$ | $6.12_{5.31}^{6.95}$ | $5.99_{4.93}^{6.82}$ |
| $L_{\text{TIR}}^{U=100}$ | log $L_\odot$ | $5.92_{5.27}^{6.71}$ | $6.62_{5.54}^{8.51}$ | $6.13_{5.53}^{6.97}$ | $5.75_{5.07}^{6.42}$ | $8.01_{7.08}^{9.08}$ | $9.08_{8.5}^{9.8}$ | $7.69_{6.97}^{8.35}$ | $7.53_{6.59}^{8.3}$ |
| $L_{\text{TIR}}^{U=1000}$ | log $L_\odot$ | $5.66_{5.01}^{6.45}$ | $6.36_{5.29}^{8.26}$ | $5.88_{5.27}^{6.71}$ | $5.49_{4.82}^{6.16}$ | $7.75_{6.83}^{8.82}$ | $8.82_{8.24}^{9.54}$ | $7.43_{6.71}^{8.09}$ | $7.27_{6.33}^{8.04}$ |
| $\frac{L_{\text{TIR}}^{U=1000}}{L_{\text{bol}}}$ | ... | $1.58_{0.7}^{3.28}$ | $1.76_{0.89}^{4.19}$ | $1.63_{0.74}^{3.49}$ | $1.52_{0.66}^{3.14}$ | $0.97_{0.6}^{1.88}$ | $1.62_{1.09}^{2.19}$ | $0.84_{0.55}^{1.34}$ | $0.84_{0.52}^{1.33}$ |
| $A_{\text{H}\alpha}$ | mag | $0.63_{0.34}^{1.05}$ | $0.86_{0.43}^{1.38}$ | $0.72_{0.54}^{1.16}$ | $0.58_{0.29}^{0.95}$ | $1.06_{0.66}^{1.6}$ | $1.53_{0.91}^{2.01}$ | $0.94_{0.62}^{1.32}$ | $0.87_{0.58}^{1.32}$ |
| $A_V$ | mag | $0.8_{0.43}^{1.33}$ | $1.09_{0.54}^{1.75}$ | $0.91_{0.54}^{1.47}$ | $0.73_{0.37}^{1.2}$ | $1.34_{0.84}^{2.02}$ | $1.94_{1.15}^{2.54}$ | $1.19_{0.78}^{1.67}$ | $1.1_{0.73}^{1.67}$ |
| $\tau_{21\mu m}$ | ... | $0.03_{0.02}^{0.05}$ | $0.04_{0.02}^{0.06}$ | $0.03_{0.02}^{0.05}$ | $0.03_{0.01}^{0.04}$ | $0.05_{0.03}^{0.07}$ | $0.07_{0.04}^{0.09}$ | $0.04_{0.03}^{0.06}$ | $0.04_{0.03}^{0.06}$ |
| $\langle \tau_{\text{UV}} \rangle$ | ... | $1.6_{0.85}^{2.64}$ | $2.17_{1.08}^{3.48}$ | $1.81_{1.07}^{2.92}$ | $1.45_{0.74}^{2.38}$ | $2.66_{1.66}^{4.02}$ | $3.85_{2.29}^{5.06}$ | $2.37_{1.56}^{3.33}$ | $2.18_{1.46}^{3.32}$ |
| $P_{\text{Reprocessed}}^{\text{Rad}}$ | log K cm$^{-3}$ | $2.96_{2.29}^{3.47}$ | $3.07_{2.5}^{3.73}$ | $3.05_{2.42}^{3.55}$ | $2.89_{2.21}^{3.39}$ | $3.2_{2.66}^{4.01}$ | $4.02_{3.6}^{4.35}$ | $3.0_{2.57}^{3.41}$ | $2.89_{2.47}^{3.36}$ |
| $P_{\text{Direct}}^{\text{Rad}}$ | log K cm$^{-3}$ | $4.23_{3.85}^{4.53}$ | $4.21_{3.66}^{4.74}$ | $4.27_{3.84}^{4.58}$ | $4.21_{3.84}^{4.49}$ | $4.55_{4.2}^{5.06}$ | $5.08_{4.68}^{5.24}$ | $4.45_{4.16}^{4.72}$ | $4.4_{4.1}^{4.66}$ |
| $P_{\text{Ion}}^{\text{Rad}}$ | log K cm$^{-3}$ | $3.74_{3.47}^{3.97}$ | $3.69_{3.29}^{4.16}$ | $3.76_{3.49}^{4.01}$ | $3.73_{3.46}^{3.94}$ | $3.97_{3.65}^{4.49}$ | $4.49_{4.09}^{4.63}$ | $3.87_{3.6}^{4.14}$ | $3.83_{3.56}^{4.07}$ |
| $P_{\text{Therm}}$ | log K cm$^{-3}$ | $5.6_{5.13}^{5.69}$ | $5.19_{5.0}^{5.62}$ | $5.56_{5.21}^{5.69}$ | $5.61_{5.06}^{5.69}$ | $5.11_{4.94}^{5.35}$ | $5.21_{5.1}^{5.35}$ | $5.07_{4.93}^{5.27}$ | $5.07_{4.9}^{5.41}$ |
| $f_{\text{trap}}$ | ... | $0.05_{0.02}^{0.11}$ | $0.07_{0.04}^{0.14}$ | $0.06_{0.03}^{0.13}$ | $0.05_{0.02}^{0.1}$ | $0.05_{0.02}^{0.09}$ | $0.09_{0.06}^{0.14}$ | $0.04_{0.02}^{0.06}$ | $0.03_{0.02}^{0.06}$ |
| $\frac{P_{\text{Ion}}^{\text{Rad}}}{P_{\text{Direct}}^{\text{Rad}}}$ | ... | $0.3_{0.26}^{0.42}$ | $0.27_{0.25}^{0.36}$ | $0.29_{0.25}^{0.36}$ | $0.31_{0.26}^{0.46}$ | $0.26_{0.24}^{0.29}$ | $0.24_{0.24}^{0.27}$ | $0.26_{0.25}^{0.3}$ | $0.27_{0.25}^{0.31}$ |
| $\frac{P_{\text{Therm}}}{P_{\text{Direct}}^{\text{Rad}}}$ | ... | $17.94_{9.05}^{37.42}$ | $12.39_{2.87}^{40.4}$ | $14.75_{7.5}^{29.16}$ | $20.56_{10.95}^{43.78}$ | $3.54_{1.44}^{7.19}$ | $1.44_{0.97}^{2.8}$ | $4.44_{2.76}^{7.68}$ | $4.88_{2.97}^{9.89}$ |

**Note.** Summary of H II region properties: circular radius $R_{\text{circ}}$, attenuation-corrected $L_{\text{H}\alpha}^{\text{corr}}$, bolometric luminosity $L_{\text{bol}}$ (Sections 2.1 and 2.3), JWST-MIRI filter luminosities at F770W, F1000W, F1130W, and F2100W (Section 3.1), total IR luminosity $L_{\text{TIR}}$ at $U=100$ and $U=1000$ (Appendix B), the ratio of total IR-to-bolometric luminosity $L_{\text{TIR}}^{U=1000}/L_{\text{bol}}$, $A_{\text{H}\alpha}$, V-band attenuation $A_V$, optical depth at 21 $\mu$m $\tau_{21\mu m}$ and radiation pressure-effective optical depth $\langle \tau_{\text{UV}} \rangle$ (Table 2), pressures $P_{\text{Reprocessed}}^{\text{Rad}}$ (Section 3.1), $P_{\text{Direct}}^{\text{Rad}}$ (Section 3.2), $P_{\text{Ion}}^{\text{Rad}}$ (Section 3.3), $P_{\text{Therm}}$, $f_{\text{trap}} = P_{\text{Reprocessed}}^{\text{Rad}}/P_{\text{Direct}}^{\text{Rad}}$, $P_{\text{Ion}}^{\text{Rad}}/P_{\text{Direct}}^{\text{Rad}}$, and $P_{\text{Therm}}/P_{\text{Direct}}^{\text{Rad}}$ (Section 4). The 16th–50th–84th percentile range for each quantity is presented for all regions, regions in galactic centers, spiral arms (and bars), and inter-arm regions. Finally, both unweighted and $L_{\text{H}\alpha}^{\text{corr}}$-weighted percentiles are included for comparison.

for variations in $P_{\text{Reprocessed}}^{\text{Rad}}$ among galactic environments. H II regions in galactic centers show the highest $P_{\text{Reprocessed}}^{\text{Rad}}/k_B \sim 10^4$ K cm$^{-3}$, followed by regions in spiral arms. Inter-arm regions show the lowest $P_{\text{Reprocessed}}^{\text{Rad}}/k_B \sim 800$ K cm$^{-3}$. These environmental variations are consistent with the most luminous regions having the highest $P_{\text{Reprocessed}}^{\text{Rad}}$.

These are the first empirical estimates of $P_{\text{Reprocessed}}^{\text{Rad}}$ for a large sample of H II regions outside the Local Group. We discuss that they appear consistent with previous Milky Way and Magellanic Cloud studies in Section 4. Here we emphasize that these $P_{\text{Reprocessed}}^{\text{Rad}}$ are low, due to the low IR optical depths in our regions. In fact, the total IR luminosities of our regions are comparable to their bolometric luminosities (Table 3), but the IR optical depths are $\approx 0.03$. Since $P_{\text{Reprocessed}}^{\text{Rad}} \propto \tau_{\text{IR}}$, the IR pressures are also low.

### 3.2. Direct Radiation Pressure on Dust

Similar to $P_{\text{Reprocessed}}^{\text{Rad}}$, "direct" radiation pressure $P_{\text{Direct}}^{\text{Rad}}$ on dust grains by UV and optical photons depends on the luminosity of the source powering each H II region, the amount of dust present, the absorption and scattering properties of dust grains, and the size and geometry of the region. We estimate $P_{\text{Direct}}^{\text{Rad}}$ using Equation (4), $R_{\text{circ}}$ (Section 2.3), and assuming a shell geometry ($\mathcal{K} = 1$).

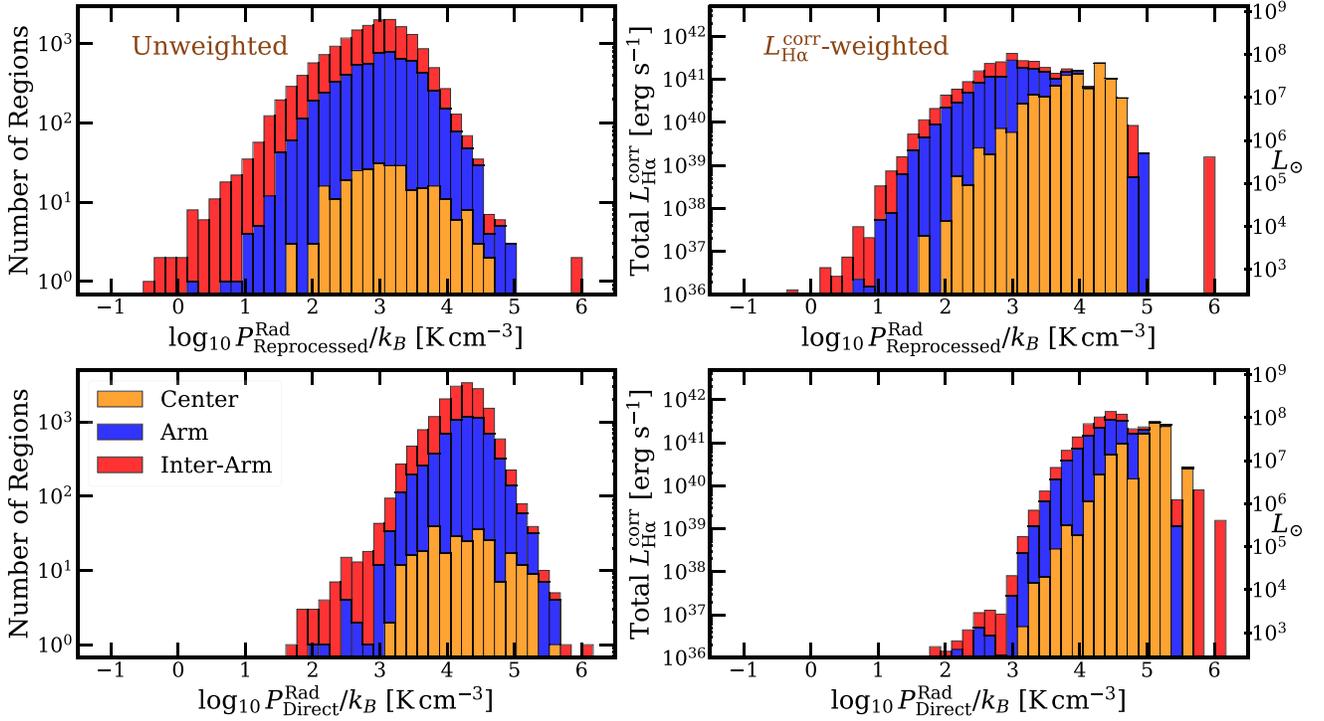

**Figure 5.** Distribution of total IR $\log P^{\rm Rad}_{\rm Reprocessed}/k_B$ (top) and UV-optical $\log P^{\rm Rad}_{\rm Direct}/k_B$ (bottom). Regions are again split by local environment—galactic centers (yellow), spiral arms and bars (blue), and inter-arm regions (red). The left panel treats each region equally, while the right panel weights regions by their $L^{\rm corr}_{\rm H\alpha}$, similar to the right panel in Figure 3.

Since young, massive stars emit most of their luminosity as UV photons, we treat the bolometric luminosity, $L_{\rm bol}$, as the relevant one. We estimate $L_{\rm bol}$ from the attenuation-corrected $L^{\rm corr}_{\rm H\alpha}$ as described in Section 2. Figure 6 and Table 3 includes the distribution of $L^{\rm corr}_{\rm H\alpha}$ and the corresponding $L_{\rm bol}$ for our sample. Our regions show median $L_{\rm bol} \approx 10^5 L_\odot$, which rises in galaxy centers to $10^6 L_\odot$. The corresponding $L^{\rm corr}_{\rm H\alpha}$-weighted median $L_{\rm bol}$ for our full sample and galaxy centers are $7 \times 10^7 L_\odot$ and $5 \times 10^8 L_\odot$.

To estimate what fraction of $L_{\rm bol}$ is attenuated and results in momentum transfer, we calculate a "radiation pressure-mean" optical depth, $\langle \tau_{\rm UV} \rangle$, following I. Blackstone & T. A. Thompson (2023). Similar to Section 3.1, we begin with $A_{\rm H\alpha}$ estimated from the Balmer decrement. Then we use the G23 extinction curve to estimate the corresponding $\tau_{\rm 150\,nm}$, the optical depth at $\lambda = 150$ nm, a characteristic FUV wavelength.

While $\tau_{\rm 150\,nm}$ offers a useful point of reference, young stellar populations emit significant energy from FUV to optical wavelengths, and the relevant optical depth for $P^{\rm Rad}_{\rm Direct}$ should reflect the convolution of the stellar SED and the wavelength-dependent dust opacity and include the effects of scattering (see above). To account for these effects, we adopt the fiducial model from I. Blackstone & T. A. Thompson (2023) and a dust-to-gas ratio of 1/100. They use a realistic grain size distribution, take into account wavelength-dependent absorption and scattering, and convolve these with model SEDs appropriate for a range of stellar population ages. In detail, we use their model to calculate the ratio between their spectrum-integrated opacity for radiation pressure $\kappa_{\rm RP}$, which is the relevant opacity for radiation pressure, and $\kappa_{\rm F,150\,nm} \propto \tau_{\rm 150\,nm}$, which we infer from our data. This yields

$$\langle \tau_{\rm UV} \rangle = 0.78 \tau_{\rm 150\,nm} \quad (8)$$

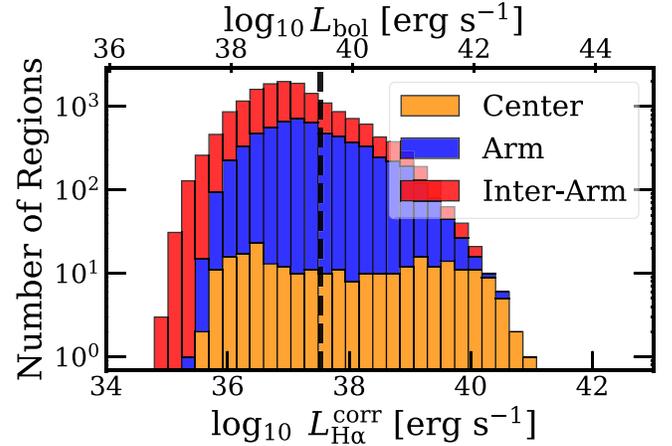

**Figure 6.** The distribution of attenuation-corrected $L^{\rm corr}_{\rm H\alpha}$. The alternate x-axis shows the corresponding $L_{\rm bol}$ estimated by assuming the average $L_{\rm bol}/L^{\rm corr}_{\rm H\alpha}$ ratio as described in Section 2.3 The vertical dashed line indicates the luminosity of a 1000 $M_\odot$ SSP at the average age of our adopted model.

for 0–1 Myr old SSP with a <10% variation across SSP ages <5 Myr (see discussion in Section 5.4). Based on this, our regions show a median $\langle \tau_{\rm UV} \rangle$ of 1.6 and $L^{\rm corr}_{\rm H\alpha}$-weighted median of 2.66, confirming that they are optically thick to starlight. This $\langle \tau_{\rm UV} \rangle$ is 2 orders of magnitude above the IR $\tau_{\rm FX}$ found in Section 3.1.

Putting together our estimates for $L_{\rm bol}$ and $\langle \tau_{\rm UV} \rangle$ yields $P^{\rm Rad}_{\rm Direct}$ for each region, as summarized in Figure 5 and Table 3. Our regions show median $P^{\rm Rad}_{\rm Direct}/k_B \approx 2 \times 10^4$ K cm$^{-3}$. H II regions in galaxy centers and spiral arms show the highest values, of the order of $10^5$ K cm$^{-3}$. These $P^{\rm Rad}_{\rm Direct}$ estimates are about an order of magnitude above $P^{\rm Rad}_{\rm Reprocessed}$, and are comparable to the ambient ISM pressure and turbulent pressure, but an order of magnitude below the thermal warm H II gas pressure $P_{\rm Therm}$ (see Section 4.3).

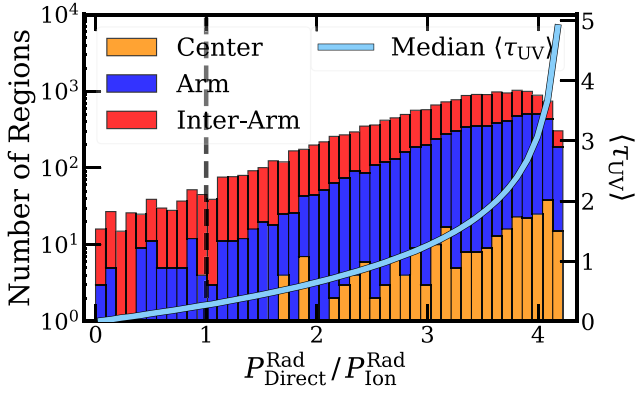

**Figure 7.** The distribution of $P_{\rm Direct}^{\rm Rad}/P_{\rm Ion}^{\rm Rad}$ for our target regions split by local galactic environment. The light-blue line shows the median value of $\langle \tau_{\rm UV} \rangle$ (right y-axis) for the regions in each bin. The vertical dashed line indicates where $P_{\rm Direct}^{\rm Rad} = P_{\rm Ion}^{\rm Rad}$. This corresponds to $\langle \tau_{\rm UV} \rangle = 0.27$.

### 3.3. Radiation Pressure from Gas Ionization

In addition to $P_{\rm Direct}^{\rm Rad}$ on dust, the photoionization of gas leads to additional radiation pressure, $P_{\rm Ion}^{\rm Rad}$. Following the approach discussed in A. T. Barnes et al. (2021), we estimate $P_{\rm Ion}^{\rm Rad}$ by considering the photon momentum contribution from the ionizing photons responsible for the observed H$\alpha$ emission.

We estimate the rate at which photons are absorbed due to H-ionization interactions, $N_{\rm abs}$, from the extinction-corrected H$\alpha$ luminosity, $L_{\rm H\alpha}^{\rm corr}$, assuming case B recombination, so that

$$N_{\rm abs} \simeq L_{\rm H\alpha}^{\rm corr}/(0.45 \times h\nu_{\rm H\alpha}), \quad (9)$$

where $h\nu_{\rm H\alpha}$ is the energy of an H$\alpha$ photon (e.g., B. T. Draine 2011b; N. Byler et al. 2017). This corresponds to a luminosity of absorbed H-ionizing photons of

$$L_{\rm Ion}^{\rm abs} = h\langle\nu_{\rm Ion}\rangle \times \frac{L_{\rm H\alpha}^{\rm corr}}{0.45 \times h\nu_{\rm H\alpha}}, \quad (10)$$

where $h\langle\nu_{\rm Ion}\rangle$ is the mean energy of an ionizing photon. In detail, $h\langle\nu_{\rm Ion}\rangle$ depends on the SED of the central powering source, which is sensitive to the age of the stellar population and potentially affected by stochastic sampling of the IMF. Lacking precise stellar mass and age estimates for individual regions, we adopt $h\langle\nu_{\rm Ion}\rangle = 18$ eV, which represents a reasonable estimate for a young stellar population with a fully sampled IMF (A. T. Barnes et al. 2021). This translates to

$$L_{\rm Ion}^{\rm abs} \approx 21 L_{\rm H\alpha}^{\rm corr}. \quad (11)$$

Using this luminosity, we compute $P_{\rm Ion}^{\rm Rad}$ following Equation (4). We adopt $\tau = \infty$ because $L_{\rm Ion}^{\rm abs}$ only refers to the ionizing photons that have been absorbed by H and led to H-ionization.

$P_{\rm Ion}^{\rm Rad}$ represents a distinct but potentially overlapping term with $P_{\rm Direct}^{\rm Rad}$. In the limit where the dust optical depth is high, $P_{\rm Direct}^{\rm Rad}$ will already account for absorption of ionizing photons because we have used the full $L_{\rm bol}$ when calculating $P_{\rm Direct}^{\rm Rad}$. However, at low depth, there will be photons that ionize the gas, but that would not have otherwise been blocked by dust.

This can be an important term, given that for a 1–4 Myr old SSP, ionizing photons ($\geqslant$13.6 eV) account for $\approx$10%–40% of the total luminosity (e.g., C. Leitherer et al. 1999). The condition where the two become comparable is

$$\begin{aligned} P_{\rm Direct}^{\rm Rad} &\lesssim P_{\rm Ion}^{\rm Rad}, \\ (1 - e^{-\langle\tau_{\rm UV}\rangle})\frac{88 \times L_{\rm H\alpha}^{\rm corr}}{4\pi R_{\rm circ}^2 c} &\lesssim \frac{21 \times L_{\rm H\alpha}^{\rm corr}}{4\pi R_{\rm circ}^2 c}, \\ \langle\tau_{\rm UV}\rangle &\lesssim 0.27. \end{aligned} \quad (12)$$

So for $\langle\tau_{\rm UV}\rangle \gtrsim 0.27$, pressure on dust will dominate over gas ionization. This corresponds to $A_V \gtrsim 0.13$ mag for our adopted extinction curve.

Figure 7 shows this ratio for our sample. Only 2% of our target regions have $\langle\tau_{\rm UV}\rangle \leqslant 0.27$, and a median $P_{\rm Ion}^{\rm Rad}/P_{\rm Direct}^{\rm Rad} \approx 0.3$. $P_{\rm Ion}$ represents an important but not dominant term in our sample. $P_{\rm Ion}^{\rm Rad}$ does dominate over $P_{\rm Direct}^{\rm Rad}$ in regions with low dust attenuation. As a result, this term will be important in low-metallicity galaxies, which are often dwarf galaxies, such as most H II regions in the relatively dust-poor Small Magellanic Cloud (SMC), which show $A_V \lesssim 0.13$ mag (L. A. Lopez et al. 2014). IC 5332 and NGC 5068, the two dwarf galaxies in our sample, indeed show the highest $P_{\rm Ion}^{\rm Rad}/P_{\rm Direct}^{\rm Rad} \approx 0.5$.

In detail, the exact $\langle\tau_{\rm UV}\rangle$ where the two terms balance depends on details of the powering stellar population (see Section 5.4). The geometric factors in Equation (4) might also differ between different pressure terms (see Section 5.2), e.g., if ionizations initially occur within a spherical H II region ($\mathcal{K} \sim 3$), while dust is confined to the neutral gas in a shell at the edge of the region ($\mathcal{K} = 1$). However, in this case, the large $P_{\rm Ion}^{\rm Rad}$ will sweep up the gas into a shell, so it won't remain spherical for long.

Note that our estimates of $P_{\rm Ion}^{\rm Rad}$ are independent of the fraction of ionizing photons that escape from a region, $f_{\rm esc}$, because we base them on $L_{\rm H\alpha}^{\rm corr}$ from the H II region itself. However, $P_{\rm Direct}^{\rm Rad}$ is sensitive to $f_{\rm esc}$, since we use $L_{\rm H\alpha}^{\rm corr}$ to estimate $L_{\rm bol}$. If $f_{\rm esc}$ is high, $L_{\rm bol}$ will be an underestimate, and so will $P_{\rm Direct}^{\rm Rad}$.

### 4. Contextualizing Radiation Pressures

#### 4.1. Comparison with Literature Measurements

We compare our estimates of $P_{\rm Reprocessed}^{\rm Rad}$ and $P_{\rm Direct}^{\rm Rad}$ to previous measurements in Figure 8. These include H II regions in the Magellanic Clouds (L. A. Lopez et al. 2014), embedded H II regions in the Milky Way's CMZ (A. T. Barnes et al. 2020), and highly obscured ultra-compact Milky Way H II regions (UCHRs; G. M. Olivier et al. 2021). These comparison data sets computed volume-averaged pressures ($\mathcal{K} = 3$, see Section 3). Therefore, we scale the literature values by one-third to be consistent with our adopted spherical shell geometry ($\mathcal{K} = 1$).

Our treatment of optical depths also differs from the literature by accounting for the IR and UV optical depths. Previous works do not include the factor of $(1 - e^{-\langle\tau_{\rm UV}\rangle})$ when estimating $P_{\rm Direct}^{\rm Rad}$, instead assuming that the full bolometric luminosity of each region is absorbed by dust grains. In our sample, this discrepancy is typically small because most H II regions are effectively optically thick to UV photons (Section 3.2), but it will be important to consider in dust-poor environments.

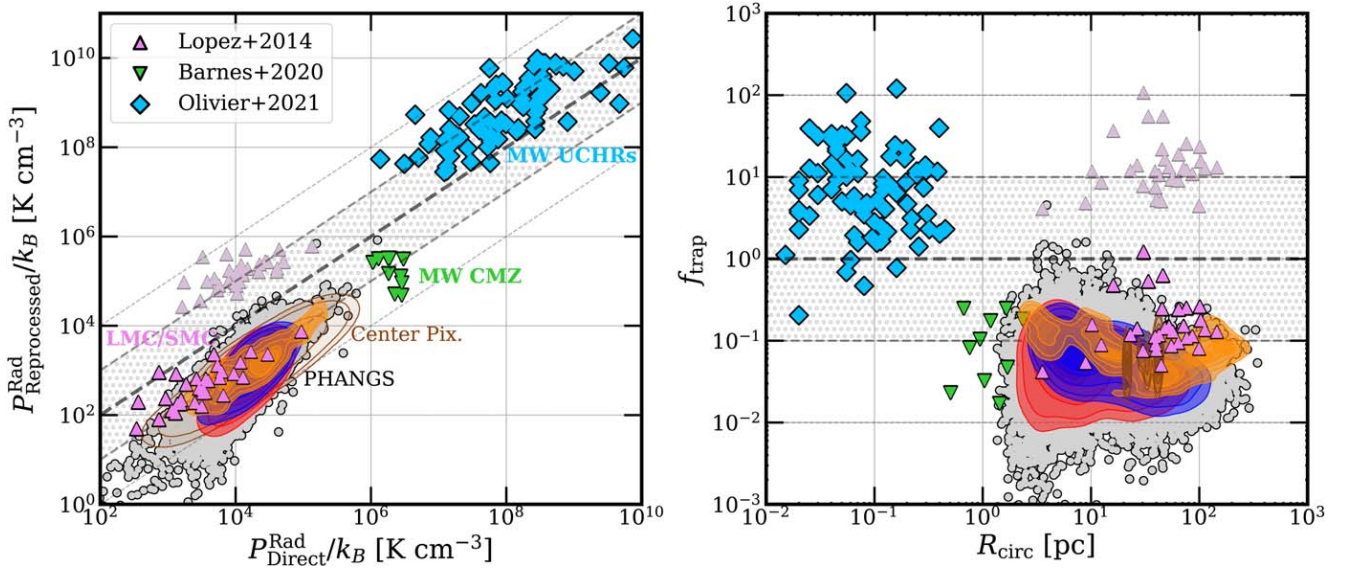

**Figure 8.** Left panel: $P_{\text{Reprocessed}}^{\text{Rad}}$ and $P_{\text{Direct}}^{\text{Rad}}$ for our ∼18,000 H II regions and literature measurements. Our individual regions are shown as gray points, and density contours represent the 16th–25th–50th–75th–84th percentile distributions colored by galactic environment—centers (yellow), spiral arms and bars (blue), and inter-arm regions (red). Brown contours show the data density for pixel-scale measurement in galactic centers (Section 2.3). Data from L. A. Lopez et al. (2014; LMC/SMC, purple triangles), G. M. Olivier et al. (2021; UCHRs, cyan diamonds), and A. T. Barnes et al. (2020; Milky Way CMZ, green triangles) is included for comparison. The original data from L. A. Lopez et al. (2014) appear as faint triangles, and bold triangles show their measurements after correcting for IR and UV optical depths (see Section 4). Right panel: similar to the left panel, now showing the variation in $f_{\text{trap}} = P_{\text{Reprocessed}}^{\text{Rad}}/P_{\text{Direct}}^{\text{Rad}}$ as a function of region radius, $R_{\text{circ}}$. The gray shaded region shows where the direct and reprocessed radiation pressure are within 1 dex of each other, around the dashed gray 1:1 line.

In addition, the works that we drew comparisons from estimated the IR-reprocessed component via $P_{\text{Reprocessed}}^{\text{Rad}} = u/3$, where $u$ is the radiation energy density in the region. While $u/3$ can reliably estimate $P_{\text{Reprocessed}}^{\text{Rad}}$ when the region is optically thick to IR photons, it becomes an unreliable estimator when $\langle\tau_{\text{IR}}\rangle \ll 1$, which is the case for most of our regions. In these optically thin regions, $P_{\text{Reprocessed}}^{\text{Rad}} = u/3$ neglects that only a small fraction of the IR radiation interacts with dust grains, and $P_{\text{Reprocessed}}^{\text{Rad}} = \langle\tau_{\text{IR}}\rangle F_{\text{IR}}/c$ is more appropriate. $P_{\text{Reprocessed}}^{\text{Rad}} = u/3$ overestimates the true value by $1/\langle\tau_{\text{IR}}\rangle$, where $\langle\tau_{\text{IR}}\rangle$ is the luminosity-weighted optical depth of IR photons.

H II regions in the LMC and SMC have low attenuations, and so we take the factor of $(1 - e^{-\langle\tau_{\text{UV}}\rangle})$ and $\langle\tau_{\text{IR}}\rangle$ into account when comparing $P_{\text{Direct}}^{\text{Rad}}$ and $P_{\text{Reprocessed}}^{\text{Rad}}$, respectively, to those measurements. L. A. Lopez et al. (2014) provided extinction estimates based on stellar reddening maps from OGLE (their Table 3), yielding median $A_V = 0.23$ mag in their LMC regions and 0.14 mag in the SMC. We use the attenuation in each region with the G23 extinction curve and compute $\langle\tau_{\text{UV}}\rangle$, and $\langle\tau_{\text{IR}}\rangle$ over $\lambda = 1$–1000 $\mu$m, following Appendix B at the $U$ provided by L. A. Lopez et al. (2014) for each region. At $U = 100$, this yields typical $\langle\tau_{\text{IR}}\rangle = 0.0039$ for the LMC and 0.0024 for SMC, firmly in the optically thin regime for IR radiation.

Figure 8 includes the original ($\langle\tau_{\text{IR}}\rangle \gg 1$) pressures from L. A. Lopez et al. (2014) as faint purple triangles, and the optical depth-corrected pressures as bold purple triangles. After correcting for $\langle\tau_{\text{IR}}\rangle$ (and $\langle\tau_{\text{UV}}\rangle$ to a smaller extent), we see good agreement between $P_{\text{Reprocessed}}^{\text{Rad}}$ and $P_{\text{Direct}}^{\text{Rad}}$ for the L. A. Lopez et al. (2014) LMC and SMC regions and H II regions in PHANGS galaxies.

The $\langle\tau_{\text{IR}}\rangle \gg 1$ assumption seems more plausible for dusty CMZ regions and UCHRs. In any case, we have no attenuation measurements for these systems. Therefore, we include them in the comparison as is, but note that obtaining estimates of $\langle\tau_{\text{IR}}\rangle$ would be valuable.

Our measurements yield ratios $P_{\text{Reprocessed}}^{\text{Rad}}/P_{\text{Direct}}^{\text{Rad}}$ consistent with those found for the Milky Way CMZ regions by A. T. Barnes et al. (2020). They do find higher pressures than even our galaxy center measurements. This is driven primarily by the small ∼1 pc sizes that they find for CMZ regions. Our HST data do not access such small scales, and we would therefore assign them larger sizes and lower pressures if they were present in our data. It seems plausible that once resolved, regions in the central regions of our target galaxies will be found to have pressures similar to the Galactic CMZ regions.

The UCHRs from G. M. Olivier et al. (2021) show much higher pressures than any of our regions. This reflects their subparsec sizes, which are 2–3 orders of magnitude smaller than PHANGS and LMC/SMC regions. In addition, G. M. Olivier et al. (2021) inferred high ISRFs in regions of high dust obscuration (where $P_{\text{Reprocessed}} \approx u/3$), which results in $P_{\text{Reprocessed}}^{\text{Rad}} > P_{\text{Direct}}^{\text{Rad}}$. The short-lived, high-pressure, IR-pressure-dominated regions in G. M. Olivier et al. (2021) are thus distinct environments from the rest of the literature measurements.

High-resolution observations of nearby galaxy centers exist in the literature and offer one prospect to bridge the gap between our measurements and subparsec Galactic regions. ALMA and VLA observations of dust and free–free emission can reach ≲1 pc resolution and resolve the regions associated with young, massive clusters (e.g., A. K. Leroy et al. 2018; K. L. Emig et al. 2020; R. C. Levy et al. 2021, 2022; E. Schinnerer et al. 2023; B. C. Whitmore et al. 2023; J. Sun et al. 2024). Measurements for star-forming regions in galaxy centers suggest high pressures and high $\langle\tau_{\text{IR}}\rangle$, implying high $P_{\text{Reprocessed}}^{\text{Rad}}/P_{\text{Direct}}^{\text{Rad}}$ (e.g., A. K. Leroy et al. 2018). Future work will synthesize these measurements in the framework used here. We discuss other next steps for region selection in Section 5.1.

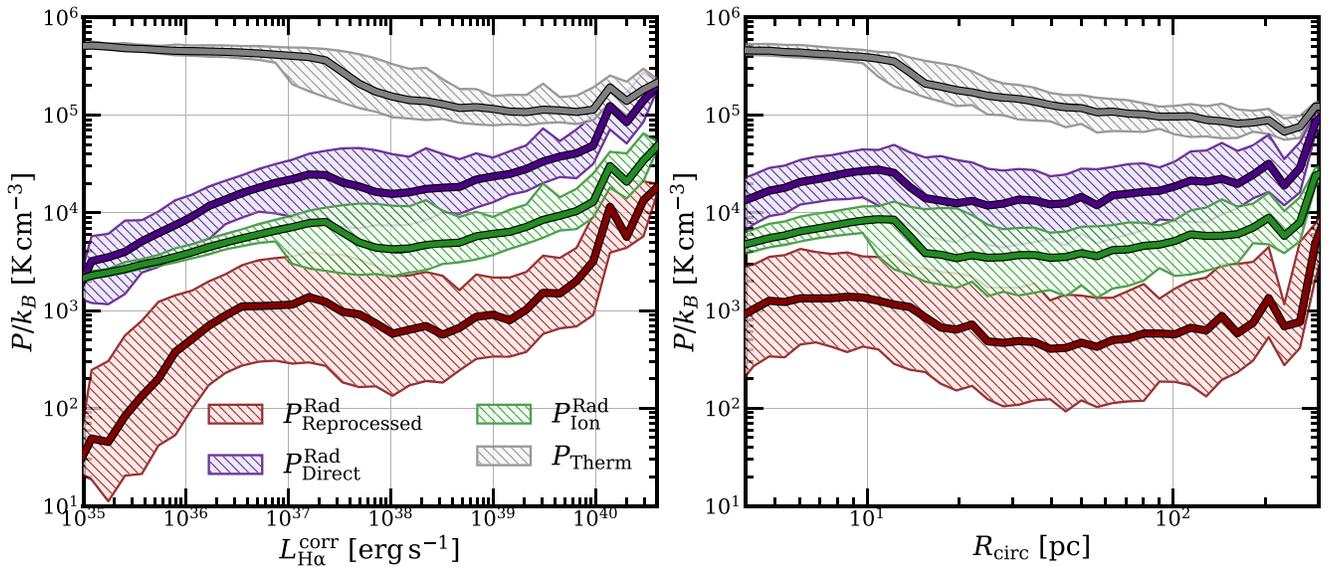

**Figure 9.** Here we compare $P_{\text{Reprocessed}}^{\text{Rad}}$ (maroon), $P_{\text{Direct}}^{\text{Rad}}$ (navy), $P_{\text{Ion}}^{\text{Rad}}$ (green), and the Strömgren $P_{\text{Therm}}$ (gray) for the full sample of H II regions. The hatched regions show the 16th–84th percentiles, and the solid lines show the median of each radiation pressure term. Variation of the pressures with $L_{\text{H}\alpha}^{\text{corr}}$ (left panel) and $R_{\text{circ}}$ (right panel) for each region is included for comparison.

Lastly, A. T. Barnes et al. (2021) estimated $P_{\text{Direct}}^{\text{Rad}}$ for a subset of our regions. Their measurements had substantial uncertainty associated with the region sizes. After accounting for their different geometric assumptions (they use $\mathcal{K} = 3$), they found mean $\log_{10} P_{\text{Direct}}^{\text{Rad}}/k_B = 5.5$ and 3.6 for their minimum and maximum sizes, representing an upper and lower bound on $P_{\text{Direct}}^{\text{Rad}}$ without the optical depth factor of $(1 - e^{-\langle\tau_{\text{UV}}\rangle})$. They focused on the most luminous subset of 6000 B. Groves et al. (2023) regions, which make our $L_{\text{H}\alpha}^{\text{corr}}$-weighted median in galaxy centers and spiral arms of $\log_{10} P_{\text{Direct}}^{\text{Rad}}/k_B = 4.45 - 5.08$ a good comparison. This lies $0.5 - 1$ dex below their upper bound for $P_{\text{Direct}}^{\text{Rad}}$, which is reasonable since the new HST-derived isophotal region sizes in the bright centers are still larger than $r_{\min}$, and trace large and luminous star-forming complexes instead of individual clusters. To illustrate, our median $R_{\text{circ}} \approx 8$ pc for all regions is a typical $r_{\min}$ for A. T. Barnes et al. (2021), while our $L_{\text{H}\alpha}^{\text{corr}}$-weighted median $R_{\text{circ}} \approx 100–170$ pc in galaxy centers and spiral arms. None of their $r_{\min}$ are $>100$ pc while this is a typical $r_{\max}$. As mentioned above, A. T. Barnes et al. (2021) also did not account for $\langle\tau_{\text{UV}}\rangle$, which would account for an additional 5%–7% decrease.

### 4.2. Comparison among Radiation Pressure Terms

Figures 9, 10, and 11 and Table 3 compare radiation pressure terms within our sample, building on Figure 8. Figure 9 shows the magnitude of each term as a function of a region's optical luminosity, $L_{\text{H}\alpha}^{\text{corr}}$, and size, $R_{\text{circ}}$. Figures 10 and 11 focus on the ratio $P_{\text{Reprocessed}}^{\text{Rad}}/P_{\text{Direct}}^{\text{Rad}}$. This ratio is also referred to as the "trapping factor,"

$$f_{\text{trap}} = \frac{P_{\text{Reprocessed}}^{\text{Rad}}}{P_{\text{Direct}}^{\text{Rad}}}. \quad (13)$$

$f_{\text{trap}}$ captures the fractional increase or "boost" in total radiation pressure due to the inclusion of reprocessed IR photons. So, $f_{\text{trap}} = 0.1$ implies that IR protons increase the total radiation pressure by 10% in addition to UV direct radiation pressure.

The figures show $P_{\text{Reprocessed}}^{\text{Rad}} < P_{\text{Ion}}^{\text{Rad}} < P_{\text{Direct}}^{\text{Rad}}$ across our sample. As we saw in Section 3.3, $P_{\text{ion}}$ is on average 0.3 times $P_{\text{Direct}}^{\text{Rad}}$, increasing to a larger fraction in the lowest-luminosity, lowest attenuation regions. For reference, the median $A_V = 0.14$ mag for the SMC mentioned in Section 4.1 implies $\langle\tau_{\text{UV}}\rangle \sim 0.28$ mag, almost exactly the condition for $P_{\text{Ion}}^{\text{Rad}} \sim P_{\text{Direct}}^{\text{Rad}}$ that we found Section 3.3. So while $P_{\text{Ion}}^{\text{Rad}}$ appears subdominant but important for our sample of massive, metal-rich galaxies, it may be dominant in the conditions similar to those found in the SMC.

Meanwhile, most massive galaxies show typical $f_{\text{trap}} = P_{\text{Reprocessed}}^{\text{Rad}}/P_{\text{Direct}}^{\text{Rad}} \approx 5\%–10\%$, a value consistent with radiation hydrodynamics simulations of giant molecular clouds (e.g., S. H. Menon et al. 2022, 2023). $f_{\text{trap}}$ is highest in galaxy centers (Table 3), higher-luminosity regions (Figure 9, left panel), and smaller regions (Figure 8, right panel), although all of these trends appear weak in our data.

Figure 10 compares the median radiation pressures and corresponding $f_{\text{trap}}$ for our 19 galaxies. Each radiation pressure correlates moderately with the global SFR (Spearman $\rho \approx 0.4–0.7$) and stellar mass ($M_*$; $\rho \approx 0.5$), and is uncorrelated with specific star formation rate (sSFR; $\rho \approx \pm 0–0.1$). While the IR-to-UV $f_{\text{trap}}$ roughly increases with SFR and $M_*$, $P_{\text{Ion}}^{\text{Rad}}/P_{\text{Direct}}^{\text{Rad}}$ decreases. As discussed in Section 3.3 and 4.1, $f_{\text{trap}}$ is lowest, and $P_{\text{Ion}}^{\text{Rad}}/P_{\text{Direct}}^{\text{Rad}}$ is highest for the two dwarf galaxies in our sample.

Finally, Figure 11 shows that $f_{\text{trap}}$ correlates with $L_{\text{F2100W}}/L_{\text{H}\alpha}^{\text{corr}}$ and $A_{\text{H}\alpha}$. We find a tight relationship between $f_{\text{trap}}$ and the luminosity ratio $L_{\text{F2100W}}/L_{\text{H}\alpha}^{\text{corr}}$, which shows $<0.2$ dex scatter. This correlation is expected. Algebraically, $f_{\text{trap}} \propto A_{\text{H}\alpha} \times L_{\text{IR}}/L_{\text{H}\alpha}^{\text{corr}}$. There is an extensive literature linking the ratio $L_{\text{H}\alpha}/L_{\text{IR}}$ to $A_{\text{H}\alpha}$ (e.g., D. Calzetti et al. 2007; F. Belfiore et al. 2023). The left panel in Figure 9 shows that $P_{\text{Direct}}^{\text{Rad}}$ and $P_{\text{Reprocessed}}^{\text{Rad}}$ scale with $L_{\text{H}\alpha}^{\text{corr}}$. The pressures also scale with IR luminosities, especially F2100W (not shown). Thus, the mid-IR-to-H$\alpha$ ratio represents our best practical predictor for $f_{\text{trap}}$, because it captures both relevant luminosities and is sensitive to the attenuation. Our best-fit bisector power-law relation, given in Figure 11, implies that $f_{\mathrm{trap}} \gtrsim 1$ when $L_{\mathrm{F2100W}}/L_{\mathrm{H}\alpha}^{\mathrm{corr}} \gtrsim 75$ and $\gtrsim 0.1$ when $L_{\mathrm{F2100W}}/L_{\mathrm{H}\alpha}^{\mathrm{corr}} \gtrsim 10$.

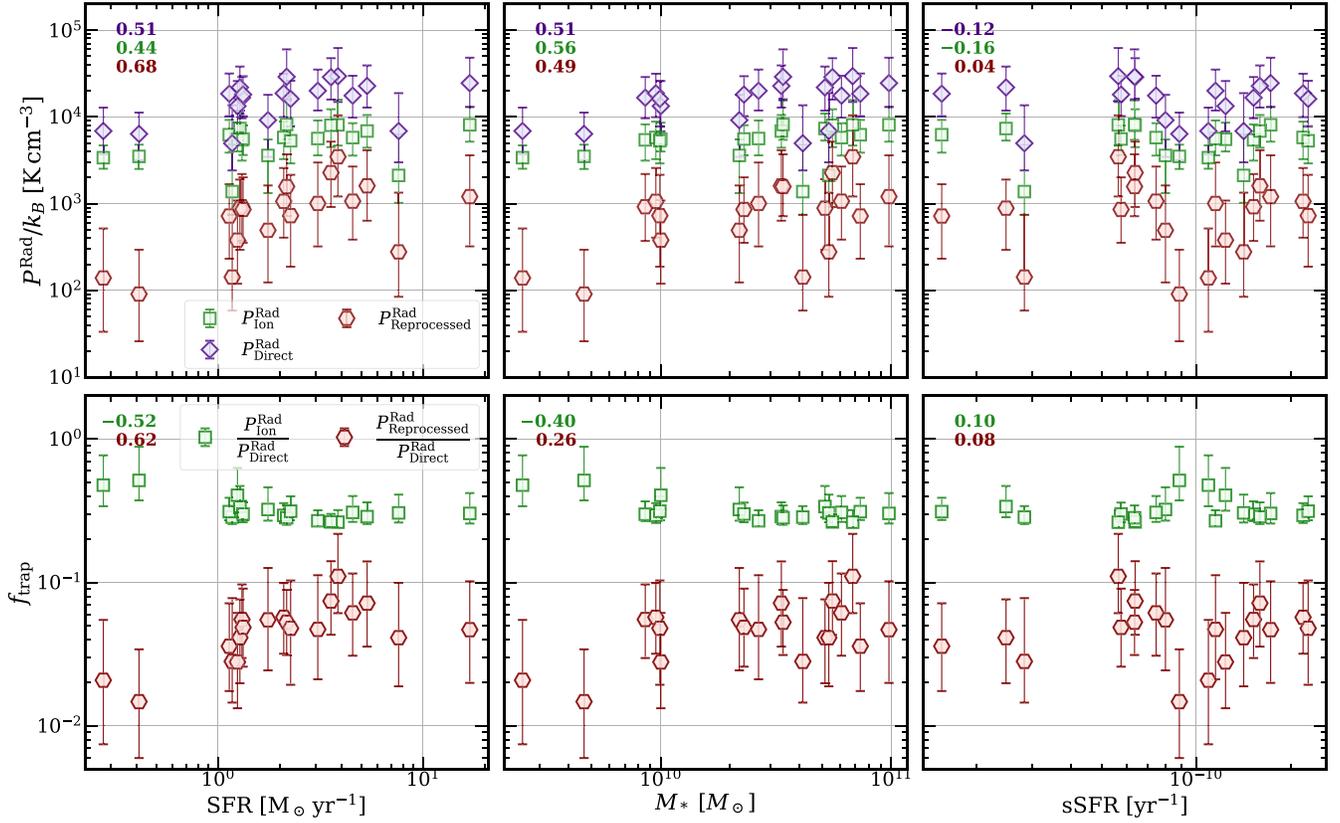

**Figure 10.** Here we compare the median $P_{\mathrm{Reprocessed}}^{\mathrm{Rad}}$ (maroon hexagons), $P_{\mathrm{Direct}}^{\mathrm{Rad}}$ (blue diamonds), and $P_{\mathrm{Ion}}^{\mathrm{Rad}}$ (green squares; top row), and $f_{\mathrm{trap}} = P_{\mathrm{Reprocessed}}^{\mathrm{Rad}}/P_{\mathrm{Direct}}^{\mathrm{Rad}}$ (maroon hexagons) and $P_{\mathrm{Ion}}^{\mathrm{Rad}}/P_{\mathrm{Direct}}^{\mathrm{Rad}}$ (green squares; bottom row) with the average star formation rate (SFR), total stellar mass ($M_*$), and specific star formation rate (sSFR) for all 19 galaxies. Each marker shows the median for each galaxy, and the error bars denote the 16th–84th percentile range in values. The Spearman rank coefficient $\rho$ for the correlation between each median and galaxy property is noted in the top left of each panel.

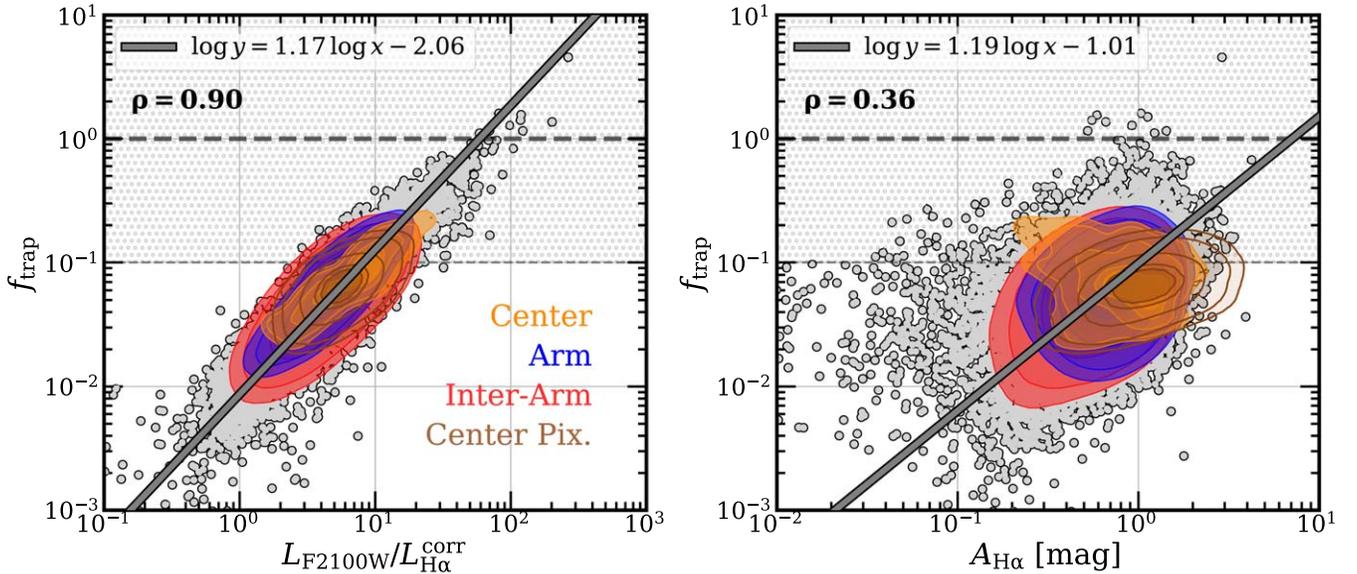

**Figure 11.** Variation in $f_{\mathrm{trap}}$ as a function of $L_{\mathrm{F2100W}}/L_{\mathrm{H}\alpha}^{\mathrm{corr}}$, which roughly captures the shape of the SED (left panel) and attenuation $A_{\mathrm{H}\alpha}$, with corresponding log–log bisector fits (solid black line). Individual regions are shown as gray background points, and density contours represent the 16th–25th–50th–75th–84th percentile distributions colored by local environment—galaxy centers (yellow), spiral arms (blue), inter-arm regions (red). Brown contours show the distribution of 0″.9 pixel-scale MUSE estimates in galaxy centers. $f_{\mathrm{trap}}$ shows a tight empirical correlation with $L_{\mathrm{F2100W}}/L_{\mathrm{H}\alpha}^{\mathrm{corr}}$, and shows much more scatter with attenuation. The gray shaded area shows the region where the direct and reprocessed radiation pressure terms are within 1 dex of each other, around the dashed gray line where $f_{\mathrm{trap}} = 1$. The Spearman rank coefficient $\rho$ for the full sample of regions is included.

In the right panel of Figure 11, $f_{\mathrm{trap}}$ shows a weaker correlation with $A_{\mathrm{H}\alpha}$, despite the expectation that $f_{\mathrm{trap}} \propto A_{\mathrm{H}\alpha}$ for regions optically thick in the UV. The two quantities do correlate as expected, but the scatter in $f_{\rm trap}$ at fixed $A_{\rm H\alpha}$ is $> \pm 0.6$ dex, and an overall Spearman rank coefficient $\rho = 0.36$. In contrast to the left panel (strong correlation, $\rho = 0.90$), the simplest explanation appears to be that $A_{\rm H\alpha}$ does not trace the $L_{\rm mid-IR}/L_{\rm H\alpha}^{\rm corr}$ ratio well ($\rho = 0.09$, not shown). Since $f_{\rm trap} \propto A_{\rm H\alpha} \times L_{\rm mid-IR}/L_{\rm H\alpha}^{\rm corr}$, scatter in that luminosity ratio at fixed $A_{\rm H\alpha}$ would weaken the observed $f_{\rm trap} - A_{\rm H\alpha}$ trend. Such scatter is present in the SFR tracer literature (e.g., see plots of $A_{\rm H\alpha}$ versus $L_{\rm H\alpha}/L_{24\mu m}$ with large scatter in A. K. Leroy et al. 2012; S. Kessler et al. 2020) but often obscured by the use of luminosity–luminosity plots.

The fact that $A_{\rm H\alpha}$ does not trace $L_{\rm mid-IR}/L_{\rm H\alpha}^{\rm corr}$ perfectly, suggests a degree of breakdown in our model, which assumes: (1) a simple spherical geometry, (2) that $L_{\rm H\alpha}^{\rm corr}$ traces the total light absorbed by dust, (3) that $L_{\rm MIRI}$ traces the relevant IR luminosity, and (4) that the VLT-MUSE $A_{\rm H\alpha}$ can be used to trace attenuation at all scales and wavelengths. We discuss the path toward improving these components, including more robust tracers of attenuation and better constraints on region geometry in Section 5.

### 4.3. Comparison to Other Pressure Terms

We also compare radiation pressure to estimates of other terms driving the expansion of H II regions. In A. T. Barnes et al. (2021), the thermal gas pressure on a shell due to the enclosed ionized gas often represented the dominant term. Assuming all H and He enclosed is singly ionized, $P_{\rm Therm} = 2k_B n_e T_e$. A. T. Barnes et al. (2021) found $P_{\rm Therm}/k_B \approx 4 \times 10^4 - 8 \times 10^5 \, \rm K \, cm^{-3}$, the range reflects uncertainty in the electron number density $n_e$. Although measurements of the ratio of the lines in the [S II] doublet offer a means of measuring $n_e$, the values were distinguishable from the low density limit only in a subset of the A. T. Barnes et al. (2021) H II regions. Also, even in cases where an [S II]-derived density could be measured, different assumptions about region size and clumpiness yielded a wide range of plausible $n_e$. The electron temperature, $T_e$, was better constrained via nitrogen auroral lines where available, and set to a representative $T_e = 8000$ K otherwise.

The new HST regions' sizes improve constraints on $n_e$, and detailed new estimates of $P_{\rm Therm}$ based on these will be presented in A. Barnes et al. (2025, in preparation). Here we provide an approximate estimate of $P_{\rm Therm}$ to compare to $P_{\rm Rad}$. We assume that the distribution of ionized gas in the H II region is well approximated by a uniform sphere (the Strömgren approximation) and adopt a representative $T_e = 8000$ K. This yields

$$\frac{P_{\rm Therm}}{k_B} = 2T_e \sqrt{\frac{3Q_0}{4\pi R_{\rm circ}^3 \alpha_B}}, \quad (14)$$

where $Q_0$ is the H-ionizing photon production rate, estimated from $L_{\rm H\alpha}^{\rm corr}$ (D. E. Osterbrock & G. J. Ferland 2006).

Figure 9 and Table 3 compare radiation pressure terms to $P_{\rm Therm}/k_B$, median value $\sim 10^{5.5} \rm K \, cm^{-3}$. Thermal gas pressure dominates over all radiation pressure terms, with $P_{\rm Therm}/P_{\rm Direct}^{\rm Rad}$ on average $\approx 18$. However, as summarized in Table 3, $P_{\rm Therm}/P_{\rm Direct}^{\rm Rad}$ becomes lower, and radiation pressure becomes more important in luminous regions where the dust attenuation is higher. In our sample, these are conditions associated with the inner, bar-fed regions of galaxies ($P_{\rm Therm}/P_{\rm Direct}^{\rm Rad} \sim 12$) or weighting by $L_{\rm H\alpha}^{\rm corr}$ ($P_{\rm Therm}/P_{\rm Direct}^{\rm Rad} \sim 3.5$). These results appear consistent with simulations (e.g., J.-G. Kim et al. 2018; A. A. Ali 2021) and models (e.g., M. R. Krumholz & C. D. Matzner 2009), which show that while photoionization heating dominates over radiation pressure for typical star-forming regions, this situation can reverse for regions powered by the most massive clusters.

Stellar winds also contribute an outward pressure, but one that is harder to constrain with observations. A. T. Barnes et al. (2021) estimated the ram pressure due to stellar winds, $P_{\rm Wind}$, by assuming the fiducial mass-loss rate, wind velocity, and mechanical luminosity in the STARBURST99 models. In this setup, $P_{\rm Wind} \propto L_{\rm bol}/R_{\rm circ}^2$, the same dependence as $P_{\rm Direct}^{\rm Rad}$. For a given size estimate, A. T. Barnes et al. (2021) found $P_{\rm wind}$ to be similar to $P_{\rm Direct}^{\rm Rad}$, typically $\sim 0.1$ dex lower. Our measurements do not contribute new constraints on this ratio, but we do note that $P_{\rm wind}$ is significantly more uncertain than $P_{\rm Direct}^{\rm Rad}$. X-ray observations do not recover the thermal energy expected from shocked hot stellar winds (L. A. Lopez et al. 2011; A. L. Rosen et al. 2014), and recent simulations suggest that rapid cooling significantly weakens the importance of shocked hot stellar winds (see, e.g., L. Lancaster et al. 2021a, 2021b).

Other terms may also contribute, for example, cosmic-ray feedback can be important on larger scales (e.g., T. A. Thompson & T. M. Heckman 2024), and the assumption that H II regions are all "pre-SN" may also be violated in some cases. We have already removed a small number of supernova remnants coincident with H II regions (see Section 2.1 and J. Li et al. 2024) from our sample. N. Mayker Chen et al. (2024) observed a minority of core collapse supernovae and a majority of stripped-envelope supernovae that occurred recently in PHANGS galaxies to overlap the B. Groves et al. (2023) regions. And in the Milky Way, LMC, and M33, supernovae are observed to occur within or near IR-bright giant H II regions (e.g., L. K. Townsley et al. 2011, 2024; S. K. Sarbadhicary et al. 2024). These composite regions will make excellent targets for case studies that examine the complex interplay of multiple generations of stellar feedback.

Finally, we have not discussed the impact of radiation pressure in region expansion. Following A. T. Barnes et al. (2021), a natural next step will be to compare the pressure of the ambient environment into which the regions expand and its balance with the outward push of stellar feedback and the restoring pull of stellar gravity. Kiloparsec-scale equilibrium ISM pressures have been estimated for our sample of galaxies by J. Sun et al. (2020, 2022) and are of the same order as $P_{\rm Therm}$ (see A. T. Barnes et al. 2021), requiring a detailed region-by-region comparison to assess their dynamical state. The stellar gravity of these regions has not been measured, but follow-up work leveraging PHANGS-JWST and PHANGS-HST imaging to constrain this term is underway.

## 5. Key Assumptions and Next Steps

### 5.1. Region Selection

Although we use HST sizes, we study regions identified as H$\alpha$ peaks in the seeing-limited ($\sim 70$ pc) VLT-MUSE maps. This tends to emphasize luminous, H$\alpha$-bright regions, and our results can be viewed as reflecting the "middle-aged" pre-supernova H II regions that contribute most of the H$\alpha$ emission in galaxies. They may miss other types of H II regions.

A particular concern is that basing our sample on B. Groves et al. (2023) excludes young, embedded regions in the earliest stages of region evolution. We expect these to be compact, with high IR-to-Hα ratios. These are likely to be the cases with the largest $\langle\tau_{IR}\rangle$ and highest $f_{trap}$, and represent an important bridge between our sample and the embedded regions studied in the Milky Way by G. M. Olivier et al. (2021) and A. T. Barnes et al. (2020). Though this phase may be short-lived, populations of these sources do appear to be present in our targets, visible in high-resolution imaging of 3.3μm PAH emission (M. Jimena Rodríguez et al. 2024) and long-wavelength, high-resolution observations of galaxy centers (E. Schinnerer et al. 2023; J. Sun et al. 2024, and see references in Section 4.1).

Star formation often occurs in crowded environments with populations of different ages near one another, e.g., in spiral arms, bar ends, and starburst galaxy centers. Such regions tend to be overshadowed by nearby larger, Hα-bright sources. Galactic and Magellanic Cloud observations often show such multigeneration star formation with exposed H II regions near mid-IR bright embedded sources, e.g., in regions like 30 Doradus (e.g., M. Cignoni et al. 2015) and Carina (e.g., K. V. Getman et al. 2014; M. S. Povich et al. 2019). This juxtaposition makes selecting small, IR-bright, relatively Hα-faint regions from seeing-limited Hα maps challenging. In Sections 2.1 and 4.1, we have already highlighted that this situation is particularly severe in Galactic centers, where B. Groves et al. (2023) assigned whole galactic centers to one or a few H II regions.

Given this, a clear next step will be to deliberately select compact embedded sources, particularly in galaxy centers. This may require modifying our approach, not just our selection. The spectroscopic data from MUSE that underpin our extinction measurements remain accessible only at ~0″.9 scales. Meanwhile, additional high-resolution long-wavelength observations, e.g., with ALMA or the VLA, will be needed to resolve the massive, highly obscured young clusters in galaxy centers.

### 5.2. Region Geometry

The relative geometry of dust, gas, and stars represents an important uncertainty in our calculations. We adopt the simplifying assumption that a single $R_{circ}$ and a shell geometry ($\mathcal{K} = 1$, Equation (4)) describes all thee radiation pressure terms. But these geometric factors are uncertain, with different literature studies assuming values that vary by a factor of ~3. Beyond just uncertainty in the appropriate overall geometry to assume, different feedback-related pressures may occur in different locations within the region. This will lead to different $\mathcal{K}$ and even different relevant region sizes for each term. For example, reprocessed radiation pressure may act outside the inner shells where direct radiation pressure acts, so that both have $\mathcal{K} = 1$ but distinct radii. Meanwhile, if the ionized gas in a region is volume-filling, then $\mathcal{K} \sim 3$ may be appropriate for the radiation pressure due to ionizations, while the pressure on dust might be better described by $\mathcal{K} = 1$. Finally, the thickness of dusty shells, and smaller high-attenuation clumps of dust embedded within a larger shell structure that we fail to resolve can contribute additional reprocessed radiation pressure. This may lead to a spatial disconnect between the optical (direct) and IR (reprocessed) components of $f_{trap}$ on smaller scales.

Fortunately, prospects for next steps here are clear. Current or incoming JWST observations will capture near-IR recombination line emission, e.g., Paschen α or Brackett α imaging, and high-resolution dust emission that can be used to estimate the extinction on small scales and resolve the geometry of individual regions (e.g., A. Pedrini et al. 2024). This will allow for detailed measurements of H II region substructure. These, combined with simulations that capture realistic structure (e.g., J.-G. Kim et al. 2018; S. H. Menon et al. 2022, 2023) and more panchromatic studies of Local Group H II regions should combine to better constrain the appropriate geometry to estimate feedback terms.

### 5.3. Optical Depth and Extinction Curve

We adopt the K. D. Gordon et al. (2023) extinction curves with a typical Milky Way $R_V = 3.1$. This $R_V$ appears appropriate for some photodissociation regions (e.g., A. Abergel et al. 2024), but not all of them (e.g., K. P. M. Blagrave et al. 2007), and high $R_V \approx 4$ have been observed toward cold molecular clouds (e.g., L. A. Valencic et al. 2004; E. L. Fitzpatrick & D. Massa 2007). $R_V$ also varies moderately from galaxy to galaxy (see review in S. Salim & D. Narayanan 2020) and exhibits both small- (cloud) and large- (kiloparsec) scale spatial variations in the Milky Way (e.g., R. Zhang et al. 2023; X. Zhang et al. 2024). We assess the impact of $R_V$ variations in Appendix C. Increasing $R_V$ from 3.1 to 4 decreases $P_{Direct}^{Rad}$ by $\lesssim 20\%$, and increases $P_{Reprocessed}^{Rad}$ by $\lesssim 20\%$, resulting in a 1.5 times higher $f_{trap}$.

Given a predictive model for $R_V$, one could adjust the G23 (or subsequent) curves region by region to reflect $R_V$ variations. There are good prospects for such a model, with JWST now regularly mapping multiple recombination lines, extensive optical spectral mapping of Milky Way and Magellanic Cloud H II regions underway (Sloan Digital Sky Survey Local Volume Mapper; N. Drory et al. 2024; K. Kreckel et al. 2024), and new UV facilities on the horizon (e.g., UVEX; S. R. Kulkarni et al. 2021).

Moreover, the K. D. Gordon et al. (2023) extinction curves only extend to ~30 μm. Beyond this, we assume $\tau \propto \lambda^\beta$ with $\beta = -2$, but this could plausibly vary (B. T. Draine 2011b). Given the ~67% contribution to IR pressure at $\lambda \gtrsim 30\mu m$ (see Figure 15 in Appendix B), this could represent an important uncertainty. Prospects for short-term improvement in understanding this term are less clear.

A related concern is the accuracy with which our Balmer-decrement-based $A_{H\alpha}$, estimated on $\gtrsim 90$ pc scales, predicts the $\tau_{IR}$ and $\langle\tau_{UV}\rangle$ relevant to $P_{Rad}$. The VLT/MUSE observations do not reveal the structure of dust within any individual region and are not sensitive to compact, deeply embedded sources. They may include contributions from diffuse ionized gas, which has lower attenuation than bright H II regions (F. Belfiore et al. 2022; E. Emsellem et al. 2022). Each B. Groves et al. (2023) region may also blend together multiple smaller star-forming regions with different true $A_{H\alpha}$ (see Section 5.1). Due to this uncertainty, as discussed in Section 4 and Figure 11, both our data and previous studies show significant scatter about the correlation between $A_{H\alpha}$ and the mid-IR-to-Hα ratio, which may indicate consistency issues related to our assumed geometry. Despite these concerns, we emphasize that these $A_{H\alpha}$ measurements do represent the best available region-by-region attenuation estimates for any sample of galaxies. Combining higher-resolution HST and JWST recombination line mapping offers the best prospect to improve on these measurements in the near future.

**Table 4**
Columns in Value-added Catalog

| Column | Unit | Description |
|---|---|---|
| gal_name | ⋯ | Galaxy name |
| region_ID | ⋯ | Nebular region ID from B. Groves et al. (2023) |
| environment | ⋯ | Local environment following M. Querejeta et al. (2021) |
| Rcirc | pc | $R_{\rm circ}$ following A. Barnes et al. (2025, in preparation; see Appendix A) |
| A_HA6562 | mag | MUSE $A_{\rm H\alpha}$ following B. Groves et al. (2023; Section 2.1) |
| A_V | mag | $A_V$ assuming G23 extinction curve and $R_V = 3.1$ (Section 2.1) |
| tau_21micron | ⋯ | $\tau_{21\mu m}$ assuming G23 extinction curve and $R_V = 3.1$ (Section 3.1) |
| tau_UV | ⋯ | $\langle\tau_{\rm UV}\rangle$ assuming G23 extinction curve, $R_V = 3.1$, following I. Blackstone & T. A. Thompson (2023; Section 3.2) |
| L_HA6562_CORR | erg s$^{-1}$ | $L_{\rm H\alpha}^{\rm corr}$, estimated to match size of region (A. Barnes et al. 2025, in preparation; Appendix A) |
| L_bol | erg s$^{-1}$ | $L_{\rm bol}$ estimated from $L_{\rm H\alpha}$ (Section 3.2) |
| L_F770W | erg s$^{-1}$ | $L_{\rm F770W}$ estimated as $\Delta\lambda_{\rm F770W}L_{\rm F770W,\lambda}$ and to match size of region (Section 3.1; Appendix A) |
| L_F1000W | erg s$^{-1}$ | $L_{\rm F1000W}$ (as $L_{\rm F770W}$) (Section 3.1) |
| L_F1130W | erg s$^{-1}$ | $L_{\rm F1130W}$ (as $L_{\rm F770W}$) (Section 3.1) |
| L_F2100W | erg s$^{-1}$ | $L_{\rm F2100W}$ (as $L_{\rm F770W}$) (Section 3.1) |
| P_Rad_TIR_Reprocessed_kB | K cm$^{-3}$ | $P_{\rm Reprocessed}^{\rm Rad}/k_B$ (Section 3.1) |
| P_Rad_MIR_F770W_kB | K cm$^{-3}$ | $P_{\rm Reprocessed,F770W}^{\rm Rad}/k_B$ (Section 3.1) |
| P_Rad_MIR_F1000W_kB | K cm$^{-3}$ | $P_{\rm Reprocessed,F1000W}^{\rm Rad}/k_B$ (Section 3.1) |
| P_Rad_MIR_F1130W_kB | K cm$^{-3}$ | $P_{\rm Reprocessed,F1130W}^{\rm Rad}/k_B$ (Section 3.1) |
| P_Rad_MIR_F2100W_kB | K cm$^{-3}$ | $P_{\rm Reprocessed,F2100W}^{\rm Rad}/k_B$ (Section 3.1) |
| P_Rad_Direct_kB | K cm$^{-3}$ | $P_{\rm Direct}^{\rm Rad}/k_B$ (Section 3.2) |
| P_Rad_Ion_kB | K cm$^{-3}$ | $P_{\rm Ion}^{\rm Rad}/k_B$ (Section 3.3) |
| P_Therm_kB | K cm$^{-3}$ | $P_{\rm Therm}/k_B$ (Section 4.3) |

**Note.** This table complements the catalogs of B. Groves et al. (2023) and A. Barnes et al. (2025, in preparation, version 0.9) with the measurements and derived quantities needed to reproduce our results. These catalogs can be joined using the region_ID from B. Groves et al. (2023). We request that users of these catalogs cite the original measurements from B. Groves et al. (2023) and A. Barnes et al. (2025, in preparation).

(This table is available in its entirety in machine-readable form in the online article.)

### 5.4. Spectral Energy Distribution and Luminosity Estimates

$P_{\rm Rad}^{\rm Direct}$ depends on the convolution of the extinction curve with the UV to IR SED of the local stellar population, which itself depends on the age and metallicity of the stellar population. When we calculate $P_{\rm Rad}^{\rm Direct}$, we use factors from I. Blackstone & T. A. Thompson (2023) that assume an SED of a typical young cluster. A clear next step will be to measure the region-by-region SEDs, corrected for extinction, and use these to refine the calculations.

Efforts are underway to measure and model the SEDs of stellar associations and clusters in PHANGS-HST (K. L. Larson et al. 2023; D. Maschmann et al. 2024). F. Scheuermann et al. (2023) and A. Barnes et al. (2025, in preparation) cross-matched HST-based stellar cluster and association catalogs to the B. Groves et al. (2023) nebular catalogs. However, F. Scheuermann et al. (2023) found that only ∼40% of the B. Groves et al. (2023) regions have overlapping associations. This reflects that many of the B. Groves et al. (2023) regions correspond to low-mass, $\lesssim 10^3 M_\odot$, stellar populations (see Figure 6). Even when there is association-region overlap, cross-matching is not straightforward because multiple associations often overlap one H II region (F. Scheuermann et al. 2023). Observations of such multigeneration stellar populations are common in Milky Way and Magellanic Cloud complexes (e.g., K. V. Getman et al. 2014; M. Cignoni et al. 2015; M. S. Povich et al. 2019), and add another layer of complexity to the modeling of individual H II by breaking the simplifying assumption of a single-age SSP.

As a result of this complexity, matched SEDs with modeled ages and masses are not yet available for our target regions.

Moving forward, we expect that aperture photometry focused on each H II region will ultimately allow similar SED modeling to be applied to the PHANGS-HST clusters (J. A. Turner et al. 2021; Thilker et al. submitted). This should replace our $L_{\rm H\alpha}^{\rm corr}$-based $L_{\rm bol}$ estimates with more realistic ones, allow for checks on our adopted attenuation, and improve the accuracy of our $\langle\tau_{\rm UV}\rangle$ estimates.

In the absence of stellar population modeling, we estimate $L_{\rm bol}$ from $L_{\rm H\alpha}^{\rm corr}$. We assume a luminosity-weighted average conversion factor $L_{\rm bol}/L_{\rm H\alpha}^{\rm corr} = 88$ over the first 4 Myr of SSP evolution. For a fully sampled IMF, this ratio depends on the age of the stellar population, with commonly used values of $L_{\rm bol}/L_{\rm H\alpha}^{\rm corr}$ ranging from ≈60 to 140 (A. T. Barnes et al. 2021; E. Schinnerer & A. K. Leroy 2024; see discussion in Section 2.1). Precise ages for each region would sharpen this estimate and allow us to distinguish a massive, somewhat older (but still likely ≲6 Myr old) region with high $L_{\rm bol}$ from a lower-mass younger region.

A related but even harder-to-address concern is stochasticity. The mean $L_{\rm H\alpha}^{\rm corr}$ for our regions corresponds to typical stellar mass ∼500–6000 $M_\odot$, which is within the regime affected by stochastic sampling of the IMF (e.g., M. Fouesneau & A. Lançon 2010; R. L. da Silva et al. 2012). Broadly, we expect that this means that the median region may have somewhat higher nonionizing UV and optical emission, and so higher $L_{\rm bol}$, than we calculate assuming a fully populated IMF. Both of these concerns should ultimately be addressed by modeling the stellar populations region by region.

Table 5 illustrates the impact of varying age (but not stochasticity), on $L_{\rm bol}/L_{\rm H\alpha}^{\rm corr}$, $\langle\tau_{\rm UV}\rangle$, inferred $M_*$, and inferred $P_{\rm direct}^{\rm Rad}$ for a fully populated IMF. Between 0 and 3 Myr, $P_{\rm Direct}^{\rm Rad}$

**Table 5**
Variation in Stellar Population Ages

| Age | $\frac{L_{\rm bol}}{L_{\rm H\alpha}^{\rm corr}}$ | $\frac{\langle\tau_{\rm UV}\rangle}{\tau_{\rm 150nm}}$ | $\frac{M_*}{M_{*,\rm ZAMS}}$ | $\frac{P_{\rm Direct}^{\rm Rad}}{P_{\rm Direct,ZAMS}^{\rm Rad}}$ |
|---|---|---|---|---|
| 0.01 Myr | 56 | 0.78 | 1 | 1 |
| 1 Myr | 61 | 0.78 | 1.01 | 1.09 |
| 2 Myr | 71 | 0.75 | 1.07 | 1.23 |
| 3 Myr | 132 | 0.72 | 1.80 | 2.25 |
| 4 Myr | 161 | 0.67 | 3.48 | 2.62 |
| 6 Myr | 637 | 0.61 | 26.85 | 9.68 |

**Note.** Typical uncertainty from varying the age of the SSP on the $L_{\rm H\alpha}^{\rm corr}$-to-$L_{\rm bol}$ conversion factor, radiation pressure-effective UV optical depth $\langle\tau_{\rm UV}\rangle$ (from I. Blackstone & T. A. Thompson 2023), inferred stellar mass relative to zero-age main sequence (ZAMS; $M_*/M_{*,\rm ZAMS}$, essentially $L_{\rm H\alpha,ZAMS}^{\rm corr}/L_{\rm H\alpha}^{\rm corr}$), and inferred direct radiation pressure relative to ZAMS ($P_{\rm Direct}^{\rm Rad}/P_{\rm Direct,ZAMS}^{\rm Rad}$).

changes by a factor of $\approx 2$ (as $L_{\rm bol}/L_{\rm H\alpha}^{\rm corr}$ rises), which is a reasonable uncertainty for any given region. This is of the same order as the uncertainty due to geometric considerations, but the age uncertainties are more likely to average out over our large sample.

Similarly, the shape of the IR SED incident on the dust is key to $P_{\rm reprocessed}^{\rm Rad}$. In Appendix B we parameterize this in terms of the intensity of the ISRF heating the dust, $U$. As shown there, these factors have a large effect on the bolometric correction to translate mid-IR to total IR luminosity. At higher $U$, a larger fraction of the total flux moves into the mid-IR as the large-grain blackbody peak shifts from the far-IR to shorter wavelengths.

On the other hand, $P_{\rm Reprocessed}^{\rm Rad}$, which depends on the optical depth-weighted intensity $\xi_\lambda$ (see Appendix B), is less dependent on changes in $U$ and is more robust to variations in the IR SED shape. As shown in Appendix B, this factor does vary, but because the optical depth drops rapidly at long wavelengths, the linear dependence on optical depth tends to offset changes in the luminosity correction from the mid-IR to TIR. The result is that the mid-IR to TIR *pressure* conversion factor $f_{\rm MIRI}^{\rm TIR}$ varies by only $\lesssim 20\%$ over the range of $U \sim 10$–$10^5$, and $\lesssim 15\%$ over $U \sim 10$–$10^4$ expected for typical H II regions.

While these calculations give us confidence in our basic approach for calculating $P_{\rm Reprocessed}^{\rm Rad}$, more empirical estimates of the IR SEDs of individual H II regions will help better constrain $U$ and hence $f_{\rm MIRI}^{\rm TIR}$. Unfortunately, it remains difficult to obtain far-IR photometry at the required high physical resolution beyond the Local Group. From studies that do exist, typical estimates of $U$ in star-forming regions in the Milky Way, LMC, SMC, M31, and M33 using far-IR maps from the Herschel Space Observatory range from $\sim 5$ to 500 (e.g., L. A. Lopez et al. 2014; J. Chastenet et al. 2019; D. Utomo et al. 2019; A. T. Barnes et al. 2020). Based on this, we consider an average $f_{\rm MIRI}^{\rm TIR}$ by varying $U$ between 1 and $10^4$, which varies $f_{\rm MIRI}^{\rm TIR}$ by $<15\%$. An improved understanding of the full IR SEDs and IR radiation field in star-forming regions remains an open topic.

Finally, some fraction of photons with $h\nu > 13.6$ eV may be absorbed by dust prior to ionizing H and producing recombination line emission. For example, B. A. Binder & M. S. Povich (2018) reported that on average, $\sim 30\%$ of Lyman continuum photons are absorbed by dust in massive IR-bright Milky Way H II regions. Without additional constraints on the fraction of H-ionizing photons absorbed in this way for our sample, we rely on $L_{\rm H\alpha}^{\rm corr}$ to infer the underlying $L_{\rm bol}$, and $A_{\rm H\alpha}$ to estimate $\tau(\lambda)$. This implies a $\lesssim 30\%$ underestimate of $L_{\rm bol}$, $\langle\tau_{\rm UV}\rangle$, and $\langle\tau_{\rm IR}\rangle$, and hence a $\lesssim 30\%$ underestimate of both $P_{\rm Direct}^{\rm Rad}$ and $P_{\rm Reprocessed}^{\rm Rad}$, but not $P_{\rm Ion}^{\rm Rad}$.

## 6. Summary

Leveraging JWST mid-IR data, VLT/MUSE optical spectral mapping, and HST measurements of H II region sizes, we estimate the strength of radiation pressure in $\sim 18,000$ H II regions across 19 nearby star-forming galaxies. These represent the first direct estimates of "reprocessed" infrared radiation pressure for a large set of H II regions outside the Local Group.

We present a new method in which the IR radiation field is estimated from high-resolution JWST mid-IR imaging, the IR extinction is extrapolated from ground-based optical nebular line ratios, and an empirically calibrated extinction curve, and region sizes are constrained by space telescope imaging data (Section 3). This combination of data, newly available by aligning JWST, HST, and VLT/MUSE, yields direct constraints on the strength of IR-reprocessed radiation pressure as well as the "direct" UV-optical pressure on dust grains and radiation pressure due to ionizations. We find:

1. The 7–21 $\mu$m range covered by JWST-MIRI captures $\approx 30\%$ of the total IR radiation pressure (though only $\sim 10\%$ of the total IR luminosity), with relatively weak dependence on the details of the radiation field (Appendix B).

2. Our full sample of H II regions has median $A_{\rm H\alpha} \approx 0.6$ mag, and a corresponding mid-IR optical depth at 21 $\mu$m of $\tau_{21\mu m} \approx 0.03$ (Section 3.1) and a typical UV optical depth at 150 nm of $\tau_{\rm 150nm} \approx 2$.

3. Across our whole sample, H II regions show median $P_{\rm Reprocessed}^{\rm Rad}/k_{\rm B} \sim 300$ K cm$^{-3}$ rising to $2 \times 10^3$ K cm$^{-3}$ when we weight by luminosity. We find higher values in galactic centers where typical $P_{\rm Reprocessed}^{\rm Rad}$ is $10^3$ K cm$^{-3}$, or $10^4$ K cm$^{-3}$ when weighting by luminosity (Section 3.1).

4. These same regions show median "direct" UV and optical radiation pressure on dust of $7 \times 10^3$ K cm$^{-3}$ and luminosity-weighted median of $3 \times 10^4$ K cm$^{-3}$. Regions in galactic centers show higher median $P_{\rm Direct}^{\rm Rad}/k_{\rm B} \sim 2 \times 10^4$ K cm$^{-3}$ and weighted by luminosity $10^5$ K cm$^{-3}$ (Section 3.2).

5. The radiation pressure exerted by H ionizing photons as they ionize gas (Section 3.3) contributes $\sim 30\%$ of the total direct optical and UV radiation pressure on average, which rises to $\sim 50\%$ in the dwarf galaxies in our sample. This term will dominate for $A_V \lesssim 0.13$ mag, a condition only met by $<2\%$ of our regions but satisfied in an average SMC H II region.

6. The ratio of $P_{\rm Reprocessed}^{\rm Rad}/P_{\rm Direct}^{\rm Rad}$ is relatively consistent across our sample, with $P_{\rm Reprocessed}^{\rm Rad}$ typically 5%–10% of the $P_{\rm Direct}^{\rm Rad}$ (Section 4, Table 3).

7. Comparing to approximate estimates of the thermal gas pressure, $P_{\rm Therm}$, the combined reprocessed, direct, and ionization radiation pressures appear subdominant (Section 4.3).

THE ASTROPHYSICAL JOURNAL, 982:140 (24pp), 2025 April 1

We emphasize that our analysis focuses on large, Hα bright regions, which represent most of the Hα emission in the PHANGS galaxies (e.g., F. Belfiore et al. 2022). Our sample does not include embedded clusters in the earliest stages of H II region evolution, where radiation pressure is expected to dominate the feedback budget. Because we base our analysis on the B. Groves et al. (2023) catalog, we do not include small, embedded clusters where $P_{\text{Reprocessed}}^{\text{Rad}}$ may dominate locally. We do expect such regions to be present in our galaxies, based on studies of 3.3 μm-bright sources (M. Jimena Rodríguez et al. 2024) and galactic centers (E. Schinnerer et al. 2023; B. C. Whitmore et al. 2023; J. Sun et al. 2024). Based on Milky Way studies, often the embedded, IR-bright sources can be near a more evolved, Hα-bright region. Therefore, a key next step will be source selection that captures the most embedded, youngest targets (e.g., H. Hassani et al. 2023; M. J. Rodrìguez et al. 2023), and local, resolved constraints on the geometry and extinction of H II regions (e.g., A. Pedrini et al. 2024). Fortunately, likely improvements seem to be on the horizon for each of these issues, and we expect to see rapid progress in this area in the near future.

## Acknowledgments

We thank the anonymous referee for providing a constructive, detailed, and insightful report. This work has been carried out as part of the PHANGS collaboration. This work is based on observations made with the NASA/ESA/CSA JWST. The data were obtained from the Mikulski Archive for Space Telescopes at the Space Telescope Science Institute, which is operated by the Association of Universities for Research in Astronomy, Inc., under NASA contract NAS 5-03127 for JWST. These observations are associated with program 2107. The specific observations analyzed can be accessed via doi: 10.17909/ew88-jt15.

This work is also based on observations collected at the European Southern Observatory under ESO programmes 094. C-0623 (PI: Kreckel), 095.C-0473, 098.C-0484 (PI: Blanc), 1100.B-0651 (PHANGS-MUSE; PI: Schinnerer), as well as 094.B-0321 (MAGNUM; PI: Marconi), 099.B-0242, 0100.B-0116, 098.B-0551 (MAD; PI: Carollo), and 097.B-0640 (TIMER; PI: Gadotti).

D.P. is supported by the NSF GRFP. A.K.L., D.P., S.S., and R.C. gratefully acknowledge support from NSF AST AWD 2205628, JWST-GO-02107.009-A, and JWST-GO-03707.001-A. A.K.L. also gratefully acknowledges support by a Humbolt Research Award. L.A.L. acknowledges support through the Heising-Simons Foundation grant 2022-3533. J.S. and K.S. acknowledge funding from JWST-GO- 2107.006-A. K.G. is supported by the Australian Research Council through the Discovery Early Career Researcher Award (DECRA) Fellowship (project No. DE220100766) funded by the Australian Government. K.G. is supported by the Australian Research Council Centre of Excellence for All Sky Astrophysics in 3 Dimensions (ASTRO 3D), through project No. CE170100013. M.B. gratefully acknowledges support from the ANID BASAL project FB210003 and from the FONDECYT regular grant 1211000. This work was supported by the French government through the France 2030 investment plan managed by the National Research Agency (ANR), as part of the Initiative of Excellence of Université Côte d'Azur under reference number ANR-15-IDEX-01. K.K. and J.E.M.-D. gratefully acknowledge funding from the Deutsche Forschungsgemeinschaft (DFG, German Research Foundation) in the form of an Emmy Noether Research Group (grant No. KR4598/2-1, PI Kreckel) and the European Research Council's starting grant ERC StG-101077573 ("ISM-METALS"). M.C. gratefully acknowledges funding from the DFG through an Emmy Noether Research Group (grant No. CH2137/1-1). COOL Research DAO is a Decentralized Autonomous Organization supporting research in astrophysics aimed at uncovering our cosmic origins. J.S. acknowledges support by the National Aeronautics and Space Administration (NASA) through the NASA Hubble Fellowship grant HST-HF2-51544 awarded by the Space Telescope Science Institute (STScI), which is operated by the Association of Universities for Research in Astronomy, Inc., under contract NAS 5-26555. S.C.O.G. acknowledges financial support from the European Research Council via the ERC Synergy Grant "ECOGAL" (project ID 855130) and from the Heidelberg Cluster of Excellence (EXC 2181 - 390900948) "STRUCTURES," funded by the German Excellence Strategy.

*Facilities:* JWST, VLT:Yepun, HST.

*Software:* astropy (Astropy Collaboration et al. 2013, 2018).

## Appendix A
## Luminosity and Size Scaling

### A.1. Optical Hα Luminosity and Isophotal Radii

HST detects only 7082 BPT-classified H II regions at good sign-to-noise ratio (SNR; A. Barnes et al. 2025, in preparation) out of the total 17,615 H II regions identified by MUSE (B. Groves et al. 2023) with $A_{\text{H}\alpha} > 0$ and reliable JWST coverage. On the other hand, the HST narrowband imaging produces more reliable sizes and region luminosities in the regions that are detected. In order to include the 10,533 regions detected by MUSE but not by HST in our analysis while retaining the benefits of the HST imaging, we translate the region sizes and optical luminosities measured by MUSE to estimates of the $L_{\text{H}\alpha}^{\text{HST}}$ and $R_{\text{circ}}^{\text{HST}}$ that we would expect to measure using HST for those regions. This allows us to estimate pressures for the full sample of 17,615 B. Groves et al. (2023) regions. To do this, we use the subset of 7082 regions with complete coverage (detected in MUSE, HST, and JWST-MIRI) to establish scaling relations that link MUSE measurements of Hα luminosity, $L_{\text{H}\alpha}^{\text{MUSE}}$, to the HST-measured luminosity $L_{\text{H}\alpha}^{\text{HST}}$ and size $R_{\text{circ}}^{\text{HST}}$.

Figure 12 shows a tight, almost linear correlation between $L_{\text{H}\alpha}^{\text{HST}}$ and $L_{\text{H}\alpha}^{\text{MUSE}}$. For regions detected by both telescopes, the power law

$$\log_{10} L_{\text{H}\alpha}^{\text{HST}} = 1.09 \log_{10} L_{\text{H}\alpha}^{\text{MUSE}} - 3.67 \quad \text{(A1)}$$

describes the relationship between the two luminosities. Individual regions show ≲0.25 dex scatter about this fit. Applying this relation results in a <30% decrease in $L_{\text{H}\alpha}$ over the range of luminosities covered by our sample.

We use the $L_{\text{H}\alpha}^{\text{HST}}$ measured by HST (from A. Barnes et al. 2025, in preparation) for the 7082 regions detected in HST, and use Equation (A1) to estimate $L_{\text{H}\alpha}^{\text{HST}}$ from $L_{\text{H}\alpha}^{\text{MUSE}}$ for the 10,533 regions without HST coverage. After adjusting the luminosity to the HST scale, we correct for the attenuation estimated from the MUSE measurement of the Balmer decrement,

$$L_{\text{H}\alpha}^{\text{HST,corr}} = L_{\text{H}\alpha}^{\text{HST}} \times 10^{A_{\text{H}\alpha}/2.5}. \quad \text{(A2)}$$

This yields the $L_{\text{H}\alpha}^{\text{corr}}$ used in Section 3. This process assumes that the attenuation measured at the MUSE resolution is also relevant at the HST resolution, which will have to be verified, e.g., with future JWST Paschen α or similar observations.

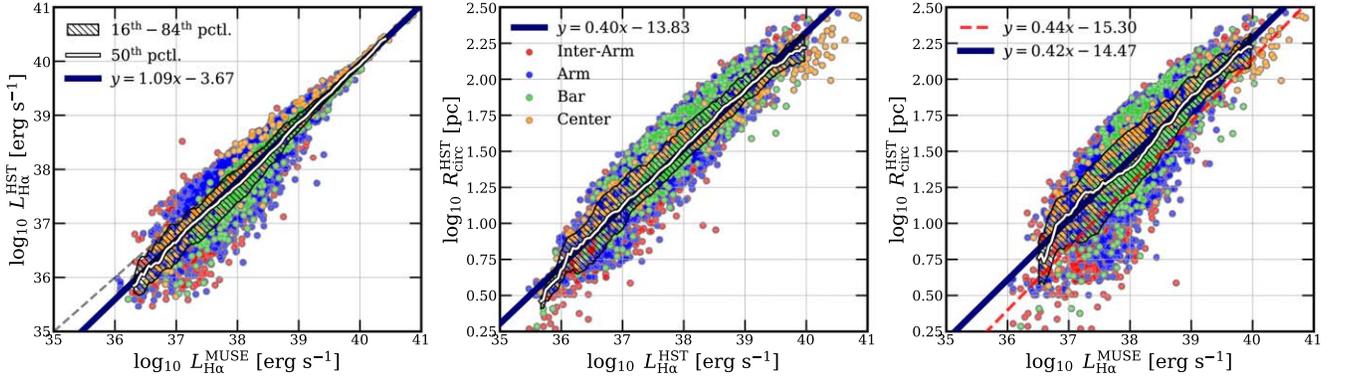

Figure 12. Correlations between (left panel) Hα luminosity in erg s$^{-1}$, as measured by MUSE and HST, (center panel) HST size and HST Hα luminosity from A. Barnes et al. (2025, in preparation), and (right panel) the closure relation between MUSE luminosities and HST sizes. All panels show the 7082 H II regions with joint HST and MUSE coverage. Individual regions are colored by their local environment—galactic centers (yellow), bars (green), spiral arms (blue), and inter-arm regions or the disk (red). The solid white line traces the 50th percentile in $y$ for each bin along the $x$-axis, and the 16th–84th percentile scatter in each bin is indicated with black hatches. Finally, the best-fit log–log correlation is shown in dark blue, which follows the 50th percentile relation very closely. The gray dashed line indicates the 1:1 line in the left panel, and the red dashed line in the right panel shows the convolution of Equations (A1) and (A3).

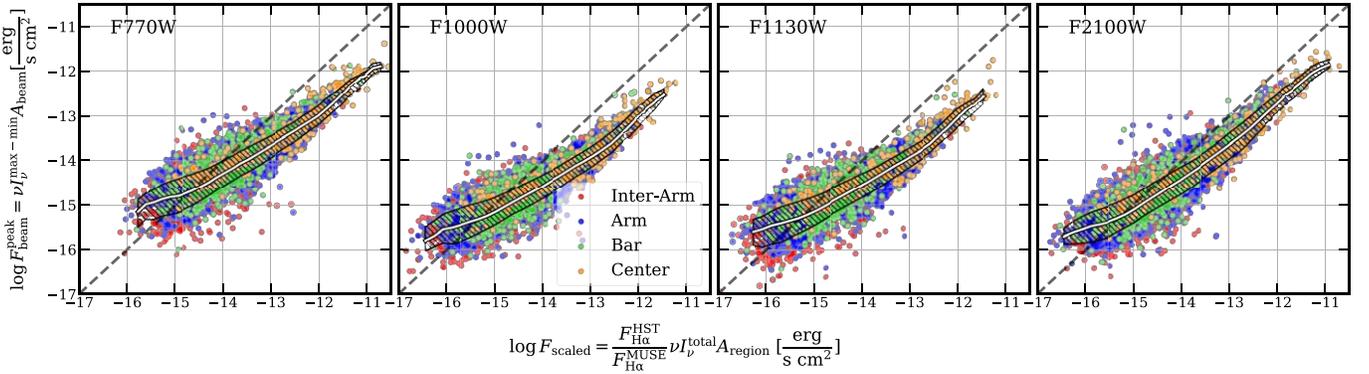

Figure 13. Comparing two methods for re-scaling MIRI fluxes from MUSE regions to HST for the ~7000 H II regions with full HST, MUSE, and JWST-MIRI coverage, colored by local environment. The $x$-axis scales MIRI fluxes directly by the same proportion that MUSE Hα fluxes are scaled relative to HST Hα fluxes. The $y$-axis calculates the total flux associated with a point source with peak intensity corresponding to the brightest pixel in each region, with a local background subtraction in each region, at 0″.9 resolution.

In addition, following A. Barnes et al. (2025, in preparation; and shown in Figure 12), we leverage the observation that the HST-measured $L_{H\alpha}^{HST}$ and circular isophotal sizes, $R_{circ}^{HST}$, show a tight power-law relationship, with ≲0.1 dex scatter,

$$\log_{10} R_{circ}^{HST} = 0.40 \log_{10} L_{H\alpha}^{HST} - 13.83. \quad (A3)$$

We use the measured $R_{circ}^{HST}$ from A. Barnes et al. (2025, in preparation) for the 7,082 HST-detected regions, and use Equation (A3) to estimate $R_{circ}^{HST}$ from $L_{H\alpha}^{HST}$ for the rest.

In practice, combining Equations (A1) and (A3) means that we use $L_{H\alpha}^{MUSE}$ to predict $R_{circ}^{HST}$. We check for closure in this relation in the right panel of Figure 12. We recover the expected correlation between $L_{H\alpha}^{MUSE}$ and $R_{circ}^{HST}$ from combining of Equations (A1) and (A3) (red dashed) within 0.25 dex.

### A.2. Mid-IR Luminosities

Since we use HST (not MUSE) Hα luminosities, for consistency, we similarly re-scale the mid-IR luminosities from JWST-MIRI for each H II region. We originally calculate these from photometry on the B. Groves et al. (2023) H II region footprints, which were derived at the VLT-MUSE resolution. The MUSE data are effectively beam-matched to our MIRI data at 0″.9. In Figure 13, we compare two methods for scaling MIRI filter fluxes from MUSE to effective HST values on the set of regions that have full HST, MUSE, and JWST coverage. The first method ($x$-axis) scales the MIRI fluxes calculated using the MUSE apertures by the MUSE-to-HST Hα scaling, as $F_{scaled} = \frac{F_{H\alpha}^{HST}}{F_{H\alpha}^{MUSE}} \nu I_\nu A_{region}$. This method of scaling remains closest to the data, and assumes the radial distribution of Hα and mid-IR fluxes on the scales of H II regions is roughly similar. We compare this to the flux of the brightest source in each region seen in the MIRI data. We calculate this from the intensity of the brightest pixel in each region after performing a local background subtraction, $I_\nu^{max-min} = I_\nu^{max} - I_\nu^{min}$, where $I_\nu^{max}$ and $I_\nu^{min}$ are the maximum and minimum pixel intensity in each region. Then we scale this intensity by the area of a Gaussian beam with FWHM $\theta = 0″.9$, $A_{beam} = \frac{\pi \theta^2}{4 \ln 2}$, to calculate the flux in a 0″.9 resolution beam, $F_{beam}^{peak} = \nu I_\nu^{max-min} A_{beam}$. This is essentially the total luminosity associated with an isolated spherical region with peak intensity corresponding to the peak intensity of each region.

For small, isolated regions, we expect the scaled HST flux to essentially capture the total flux from a point source, or $F_{beam}^{peak} \approx F_{scaled}$. Figure 13 shows that this is indeed the case, where the flux from a single point source is an excellent approximation for the mid-IR flux from fainter (smaller) regions. This also cross-validates our approach of using $F_{\text{scaled}}$ to re-scale MIRI fluxes and luminosities, since $F_{\text{scaled}}$ is better at accounting for the full flux of regions with geometries more complex than single point sources. Analogous to optical luminosities, we thus use Equation (A1) to scale all MIRI luminosities.

## Appendix B
## Mid-IR to Total IR Radiation Pressure

In the main text (Equation (7)), we used $f_{\text{MIRI}}^{\text{TIR}}$ to scale the total mid-IR pressure captured in the four MIRI filters to the total IR pressure. In this Appendix, we calculate the mid-IR-to-total IR conversion factor, $f_{\text{MIRI}}^{\text{TIR}}$, using the HD23 dust models and G23 extinction law.

The left panel of Figure 14 shows how the HD23 dust emission spectral energy distributions (SEDs) vary across 5 orders of magnitude in $U$. The HD23 dust models (as well as those of B. T. Draine et al. 2021) scale a single radiation field. Hence, for this exercise, we do not vary the SED of the radiation field, only its intensity, which is parameterized by $U$. Higher intensity $U$ can be expected from younger stellar populations and configurations where dust grains that are closer to younger clusters are exposed to higher $U$.

Using the optical depth $\tau_\lambda$ from G23, the right panel of Figure 14 shows the wavelength-distribution of pressure contribution in the optically thin limit across 5 orders of magnitude in $U$. For convenience, we define the variable $\xi_\lambda = \tau_\lambda I_\lambda$ as a proxy for the specific pressure contribution at each $\lambda$. $\xi_\lambda$ is the optical depth-weighted specific intensity, which translates to a radiation pressure per wavelength, as $P_\lambda = \xi_\lambda \Omega/c$, where $\Omega$ is the area of each MIRI pixel in steradians. We note that while changing the shape of the ISRF can be important for PAH heating, where a power-law distribution may be more realistic than assuming a single radiation field (e.g., B. T. Draine et al. 2021; D. A. Dale et al. 2023; D. Baron et al. 2024), the behavior of $f_{\text{MIRI}}^{\text{TIR}}$ remains remarkably stable across a wide range of $U$ (see discussion below), and should not significantly impact $f_{\text{MIRI}}^{\text{TIR}}$.

Integrating over the distributions in Figure 14 provides an estimate of the fraction of total-IR intensity and associated radiation pressure traced by a part or the full IR SED. We thus find the fraction of total-IR radiation pressure traced by a combination of the four MIRI filters as

$$(f_{\text{MIRI}}^{\text{TIR}})^{-1} = \frac{\sum_{\text{MIRI,FX}} \Delta\lambda_{\text{FX}} \xi_{\text{FX}}}{\int_{1\mu m}^{1000\mu m} \xi_\lambda d\lambda}, \quad (B1)$$

where $\Delta\lambda_{\text{FX}}$ is the width of each MIRI filter FX, and $\xi_{\text{FX}} = \tau_{\text{FX}} I_{\text{FX}}$. $\tau_{\text{FX}}$ is the luminosity-weighted mean G23 optical depth in each MIRI filter, which is equivalent to the optical depth at the central wavelength of each filter, since $\Delta\lambda_{\text{FX}}$ is small and $\tau_\lambda$ remains relatively flat within the width of each filter. $I_{\text{FX}}$ is the specific intensity in each MIRI filter.

Figure 15 shows the fraction of total IR intensity (top panel) and radiation pressure (middle panel) captured by several combinations of filters. Finally, the bottom panel of Figure 15 shows the corresponding conversion factors $f_i^{\text{TIR}}$ (inverting fractions from the middle panel). We present conversion factors for a number of different combinations of MIRI filters. We include versions that use the $\lambda \times I_\lambda$ or $\lambda \times \xi_\lambda$ approximation (in blue), since this approximation for energy is common in the literature. Since we prefer a running integral for better accuracy, we use the approximations $\Delta\lambda \times I_\lambda$ or $\Delta\lambda \times \xi_\lambda$ (in red), which use the filter width to more accurately capture the total intensity or pressure output of each MIRI filter. Finally, the fraction captured by the sum of all four MIRI filters, which we use to calculate total IR-reprocessed radiation pressures in Section 3.1, is indicated in solid red, and the integral over the SED in the mid-IR (1–25 $\mu$m) is in gray, for comparison.

Below $U \sim 100$, our four MIRI filters capture $\sim$15% of the TIR intensity. Above $U \sim 100$, the fraction of emission in the mid-IR steadily climbs as more thermal emission appears in the mid-IR, especially at 21 $\mu$m (e.g., see B. T. Draine & A. Li 2007). At $U \sim 10^3$, $U \sim 10^4$, and $U \sim 10^5$, the four MIRI filters capture of the order of 20%, 30%, and 40% of the TIR intensity. However, since the optical depth drops rapidly beyond the mid-IR, the contribution of long-wavelength emission to $P_{\text{Reprocessed}}^{\text{Rad}}$ is smaller than its contribution to the luminosity. While the fraction of total intensity or flux from dust emission peaks in the far-IR, the opacity-weighted intensity, which is proportional to the *force* or pressure, peaks in the mid-IR. In fact, the pressure contribution peaks at the 7.7 $\mu$m and 11.3 $\mu$m PAH features across 5 orders of magnitude in $U$, and remains fairly flat between the two PAH features, including at 10 $\mu$m, where there is less intensity but higher optical depth. In other words, the IR pressure contribution peaks in the mid-IR, between 7 and 11 $\mu$m.

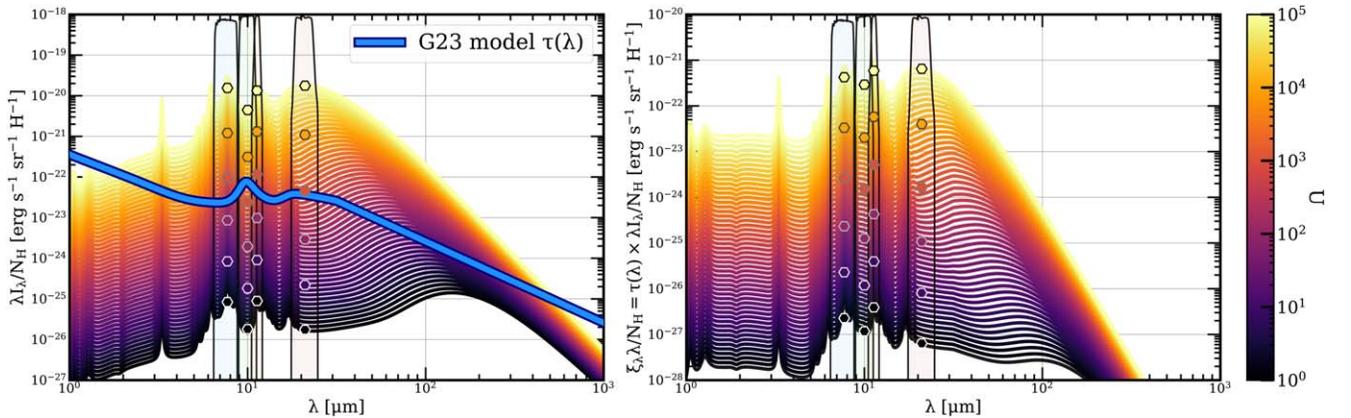

**Figure 14.** Standard HD23 dust models at different radiation fields $U$. The G23 extinction curve (arbitrary normalization, for visualization only), and the MIRI filter throughputs (arbitrary normalization) at F770W, F1000W, F1130W, and F2100W are included. The left panel shows the variation in intensity $\lambda I_\lambda$ with wavelength and $U$, showing how the SED of dust emission changes with radiation field. The right panel shows how the shape of the optical depth-weighted intensity, $\lambda \xi_\lambda = \tau(\lambda) \lambda I_\lambda$, which is a proxy for pressure contribution, varies with $U$. Hexagonal points represent the shape of the *measured* SED by showing the model SED convolved with each of the four MIRI bandpasses, at $U = 1, 10, 10^2, 10^3, 10^4$, and $10^5$.

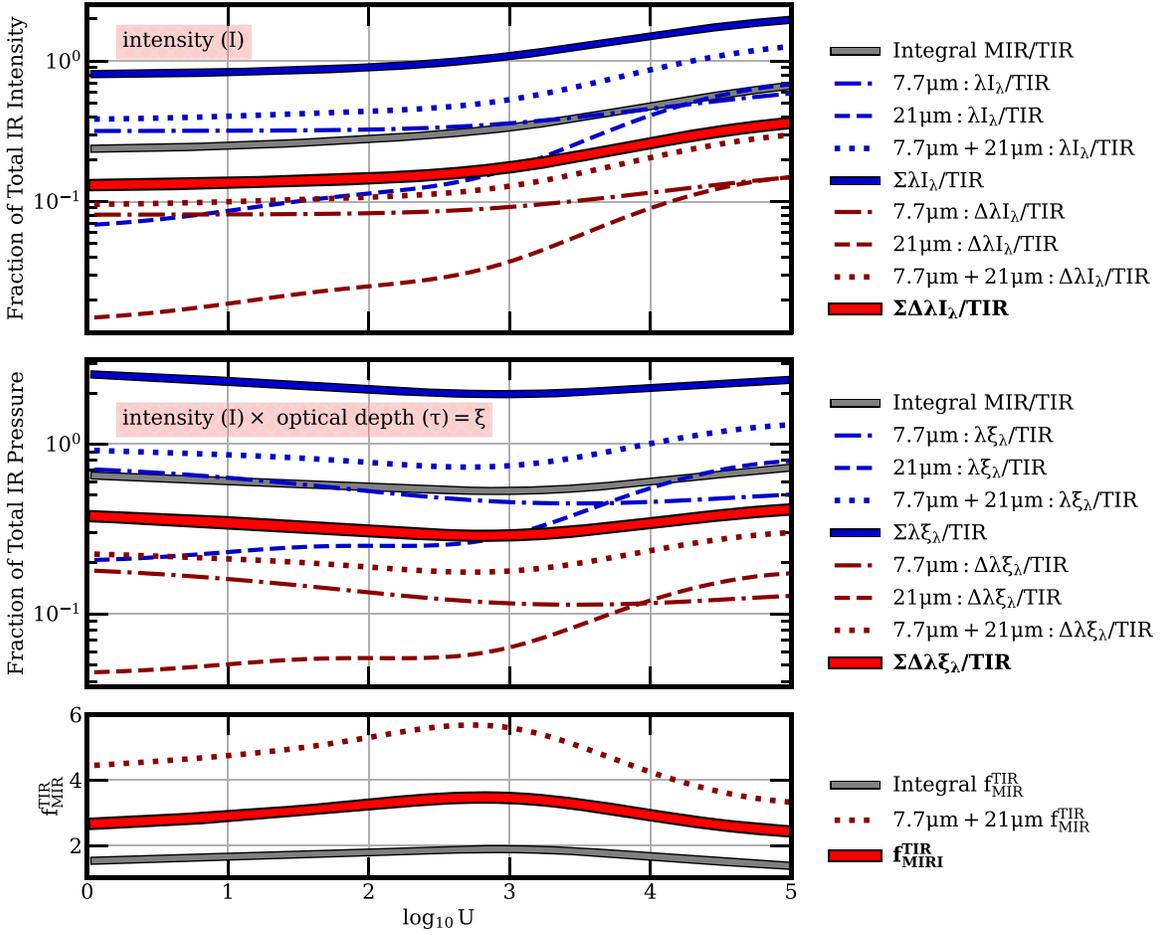

**Figure 15.** Top panel: the fraction of total IR intensity (or flux) (TIR=$\int_{1\,\mu m}^{1000\,\mu m} I_\lambda d\lambda$) captured by different approximations and linear combinations of MIRI filters at a range of $U$. First, we present the fraction of TIR flux captured by the $\lambda I_\lambda$ approximation in the 7.7 $\mu$m filter (blue dotted–dashed), 21 $\mu$m filter (blue dashed), 7.7 $\mu$m + 21 $\mu$m filters (blue dotted), and all four MIRI filters (solid blue). Next, using the filter width $\Delta\lambda$ for each MIRI filter, we present the fraction captured by $\Delta\lambda I_\lambda$ in the 7.7 $\mu$m filter (red dotted–dashed), 21 $\mu$m filter (red dashed), 7.7 $\mu$m + 21 $\mu$m filters (red dotted), and adding all four MIRI filters (solid red). Finally, we present the fraction of flux captured in the mid-IR by an integral over the model dust SED as $\int_{1\mu m}^{25\mu m} I_\lambda d\lambda / \int_{1\mu m}^{1000\mu m} I_\lambda d\lambda$ (solid gray). Middle panel: similar to the top panel, now for opacity-weighted intensity (proxy for pressure) $\xi_\lambda$ instead of intensity $I_\lambda$. Bottom panel: inverting the fractions in the middle panel provides the mid-IR to TIR correction factor $f_\xi^{\rm TIR}$ at a given $U$. The correction factor from integrating the model SED (solid gray), adding up all four MIRI filters (solid red), and 7.7 $\mu$m + 21 $\mu$m filters (red dotted) is included for comparison. The MIRI-estimated $f_\xi^{\rm TIR}$ using all four filters (solid red) is used for all $P_{\rm Reprocessed}^{\rm Rad}$ calculations.

We find that the pressure conversion factor $f_i^{\rm TIR}$ remains relatively flat at low $U$, reflecting that even at low intensity, the stochastic heating of small grains produces a steady fraction of emission in the mid-IR. This is consistent with D. A. Dale & G. Helou (2002; see their Table 2), who found that at low $U$, the mid-IR captures a steady fraction of the total IR energy budget. The fraction of TIR $P_{\rm Reprocessed}^{\rm Rad}$ captured by the four MIRI filters (solid red line) remains between 30% and 40% across the 5 orders of magnitude in $U$. A full model integral over the mid-IR range from 1–30 $\mu$m (solid gray line) captures $\sim$60%–70% of the TIR radiation pressure. In the diffuse ISM where $U \sim 1$ (D. Utomo et al. 2019), the four MIRI filters capture 13% of the total IR luminosity but 38% of the total IR pressure. At an average $U \approx 10$, which is the typical average for H II regions at 10 pc resolution in the LMC and SMC (J. Chastenet et al. 2019; D. Utomo et al. 2019), the four MIRI filters together capture 14% of the total-IR intensity or flux, but 35% of the total-IR opacity-weighted intensity; hence, $P_{\rm Reprocessed}^{\rm Rad}$. At $U \sim 300$, typical for some of the most powerful H II regions in the Milky Way CMZ (A. T. Barnes et al. 2020) or 30 Doradus in the LMC (L. A. Lopez et al. 2014), the four MIRI filters trace 35% of the total-IR flux and 32% of the total-IR pressure. Since $f_{\rm MIRI}^{\rm TIR}$ (solid red) varies between 2.7 and 3.7 over the range $U = 1$–$10^5$, we use a fiducial value of $f_{\rm MIRI}^{\rm TIR} = 3.33$ typical for H II regions where $U = 1$–$10^4$. This results in a $\lesssim$15% typical range of uncertainty in $f_{\rm MIRI}^{\rm TIR}$ over $U = 1$–$10^4$.

## Appendix C
## Radiation Pressure and Variation in $R_V$

As discussed in Section 5.3, while the radiation pressure estimates we present assume a Milky Way $R_V = 3.1$, many studies of Milky Way star-forming regions have found higher values, $R_V \approx 4$. H II regions might have $R_V > 3.1$ because smaller dust grains, including PAHs, are more prone to destruction close to strong radiation fields in young star-forming regions. Alternatively, larger dust grains may form from grain growth associated with cold clouds.

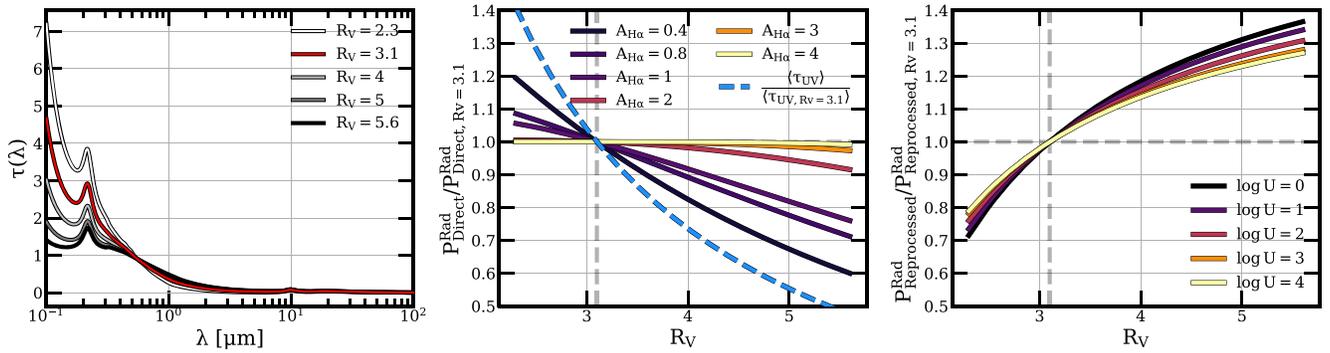

**Figure 16.** The impact of variation in $R_V$. Left panel: variation of the K. D. Gordon et al. (2023) extinction curves as a function of $R_V$. Additionally, the center panel shows the scaling of $P_{\rm Direct}^{\rm Rad}$ while the right panel shows $P_{\rm Reprocessed}^{\rm Rad}$ as a function of $R_V$ for values between 2.3 and 5.6, normalized to a our fiducial $R_V$ of 3.1. The variation in $P_{\rm Direct}^{\rm Rad}$ relative to $R_V = 3.1$ depends on the attenuation, and we show a range of $A_{\rm H\alpha}$ between 0.4 and 4. The relative variation in $P_{\rm Reprocessed}^{\rm Rad}$ depends to a smaller degree on the radiation field, shown for a range of $\log U$ between 0 and 4.

Figure 16 quantifies the systemic uncertainty in our radiation pressure estimates due to variation in $R_V$. As $R_V$ increases, the overall shape of the extinction curve gradually flattens, which results in a decrease in $\langle \tau_{\rm UV} \rangle$. This translates to a decrease in $P_{\rm Direct}^{\rm Rad} \propto (1 - e^{-\langle \tau_{\rm UV}\rangle})$. At $A_{\rm H\alpha} \gtrsim 2$, the change in $P_{\rm Direct}^{\rm Rad}$ due to varying $R_V$ is negligible: a <5% decrease at $R_V = 4$. However, for less-obscured regions where $A_{\rm H\alpha} \lesssim 1$, increasing $R_V$ from 3.1 to 4 can result in a 10%–20% lower $P_{\rm Direct}^{\rm Rad}$. Conversely, $P_{\rm Reprocessed}^{\rm Rad} \propto \langle \tau_{\rm IR} \rangle$ can increase as $R_V$ increases. Raising $R_V$ from 3.1 to 4 can result in a $\lesssim 20\%$ increase in $P_{\rm Reprocessed}^{\rm Rad}$ across a range of $U$.

This translates to a systemic increase in our estimates of $f_{\rm trap} = P_{\rm Reprocessed}^{\rm Rad} / P_{\rm Direct}^{\rm Rad}$ with increasing $R_V$. Compared to a Milky Way $R_V = 3.1$, $f_{\rm trap}$ can be up to 1.5 times higher if $R_V = 4$, and up to 2.5 times higher if $R_V = 5$.

## ORCID iDs

Debosmita Pathak ⓘ https://orcid.org/0000-0003-2721-487X
Adam K. Leroy ⓘ https://orcid.org/0000-0002-2545-1700
Todd A. Thompson ⓘ https://orcid.org/0000-0003-2377-9574
Laura A. Lopez ⓘ https://orcid.org/0000-0002-1790-3148
Ashley. T. Barnes ⓘ https://orcid.org/0000-0003-0410-4504
Daniel A. Dale ⓘ https://orcid.org/0000-0002-5782-9093
Ian Blackstone ⓘ https://orcid.org/0000-0002-7329-560X
Simon C. O. Glover ⓘ https://orcid.org/0000-0001-6708-1317
Shyam H. Menon ⓘ https://orcid.org/0000-0002-0311-2206
Jessica Sutter ⓘ https://orcid.org/0000-0002-9183-8102
Thomas G. Williams ⓘ https://orcid.org/0000-0002-0012-2142
Dalya Baron ⓘ https://orcid.org/0000-0003-4974-3481
Francesco Belfiore ⓘ https://orcid.org/0000-0002-2545-5752
Frank Bigiel ⓘ https://orcid.org/0000-0003-0166-9745
Alberto D. Bolatto ⓘ https://orcid.org/0000-0002-5480-5686
Médéric Boquien ⓘ https://orcid.org/0000-0003-0946-6176
Rupali Chandar ⓘ https://orcid.org/0000-0003-0085-4623
Mélanie Chevance ⓘ https://orcid.org/0000-0002-5635-5180
Ryan Chown ⓘ https://orcid.org/0000-0001-8241-7704
Kathryn Grasha ⓘ https://orcid.org/0000-0002-3247-5321
Brent Groves ⓘ https://orcid.org/0000-0002-9768-0246
Ralf S. Klessen ⓘ https://orcid.org/0000-0002-0560-3172
Kathryn Kreckel ⓘ https://orcid.org/0000-0001-6551-3091
Jing Li ⓘ https://orcid.org/0000-0002-4825-9367
J. Eduardo Méndez-Delgado ⓘ https://orcid.org/0000-0002-6972-6411
Erik Rosolowsky ⓘ https://orcid.org/0000-0002-5204-2259
Karin Sandstrom ⓘ https://orcid.org/0000-0002-4378-8534
Sumit K. Sarbadhicary ⓘ https://orcid.org/0000-0002-4781-7291
Jiayi Sun ⓘ https://orcid.org/0000-0003-0378-4667
Leonardo Úbeda ⓘ https://orcid.org/0000-0001-7130-2880